\documentclass[12pt,preprint2]{aastex}
\newcommand{\Msun}{\hbox{\hbox{M}$_\odot\;$}}

\newcommand{\kms}{\hbox{${\rm km}\:{\rm s}^{-1}\;$}}
\newcommand{\Msuno}{\hbox{\hbox{M}$_\odot$}}
\newcommand{\Rsuno}{\hbox{\hbox{R}$_\odot$}}
\newcommand{\Lsuno}{\hbox{\hbox{L}$_\odot$}}
\newcommand{\kmso}{\hbox{${\rm km}\:{\rm s}^{-1}$}}
\newcommand{\teff}{$T_{\rm eff}\;$}  
\newcommand{\teffo}{$T_{\rm eff}$}  
\newcommand{\logg}{$\log\;g\;$}
\newcommand{\loggl}{$\log (g/{\rm cm~s}2)$}

\begin{document}

%% LaTeX will automatically break titles if they run longer than
%% one line. However, you may use \\ to force a line break if
%% you desire.

\title{The Chemical Abundances of Tycho G in Supernova Remnant 1572}

%% Use \author, \affil, and the \and command to format
%% author and affiliation information.
%% Note that \email has replaced the old \authoremail command
%% from AASTeX v4.0. You can use \email to mark an email address
%% anywhere in the paper, not just in the front matter.
%% As in the title, use \\ to force line breaks.

\author{Jonay I. Gonz\'alez Hern\'andez\altaffilmark{1,2}, Pilar
Ruiz-Lapuente\altaffilmark{3,4}, Alexei V. Filippenko\altaffilmark{5}, 
Ryan J. Foley\altaffilmark{5}, \\ Avishay Gal-Yam\altaffilmark{6,7}, and
Joshua D. Simon\altaffilmark{6}
}

%\affil{Observatoire de Paris-Meudon, GEPI, 5 place Jules Janssen,
%92195 Meudon Cedex, France}

%\author{Pilar Ruiz Lapuente\altaffilmark{4,5}}
%\affil{National Optical Astronomy Observatories, Tucson, AZ 85719}
%\email{aastex-help@aas.org}

%\and

%\author{Alexei V. Filippenko et al.}
%\affil{Department of Astronomy, University of California, Berkeley, CA
%94720-3411.}

%% Notice that each of these authors has alternate affiliations, which
%% are identified by the \altaffilmark after each name.  Specify alternate
%% affiliation information with \altaffiltext, with one command per each
%% affiliation.

\altaffiltext{1}{Observatoire de Paris-Meudon, GEPI, 5 place Jules
Janssen, 92195 Meudon Cedex, France: Jonay.Gonzalez-Hernandez@obspm.fr .}
\altaffiltext{2}{CIFIST Marie Curie Excellence Team.}
\altaffiltext{3}{Departament d'Astronomia, Universitat de Barcelona,
Mart{\'\i} i Franqu\'es 1, E-08028 Barcelona, Spain: pilar@am.ub.es .}
\altaffiltext{4}{Max-Planck Institut f\"ur Astrophysik,
Karl-Schwarzschild-Strasse 1, D-85740 Garching, Germany:
pilar@mpa-garching.mpg.de .}
\altaffiltext{5}{Department of Astronomy, University of California, 
Berkeley, CA 94720-3411: alex@astro.berkeley.edu, 
rfoley@astro.berkeley.edu .}
\altaffiltext{6}{Department of Astronomy, California Institute of
Technology, MS 105-24, Pasadena, CA
91125:  avishay@astro.caltech.edu, jsimon@astro.caltech.edu .} 
\altaffiltext{7}{Benoziyo Center for Astrophysics, Weizmann Institute
of Science, 76100 Rehovot, Israel: avishay.gal-yam@weizmann.ac.il .}

%% Mark off your abstract in the ``abstract'' environment. In the manuscript
%% style, abstract will output a Received/Accepted line after the
%% title and affiliation information. No date will appear since the author
%% does not have this information. The dates will be filled in by the
%% editorial office after submission.

\begin{abstract}

We present an analysis of the chemical abundances of the star Tycho G
in the direction of the remnant of supernova (SN) 1572, based on Keck
high-resolution optical spectra. The stellar parameters of this star
are found to be those of a G-type subgiant with $T_{\mathrm{eff}} =
5900 \pm 100$ K, \loggl\ $ = 3.85 \pm 0.30$ dex, and $\mathrm{[Fe/H]} = -0.05
\pm 0.09$.  This determination agrees with the stellar parameters
derived for the star in a previous survey for the possible companion star 
of SN 1572 (Ruiz-Lapuente et al. 2004).  The chemical abundances follow 
the Galactic trends, except for Ni, which is overabundant relative to Fe,
$[{\rm Ni/Fe}] $ $=$ 0.16 $\pm$ 0.04. Co is slightly overabundant
(at a low significance level). These enhancements in Fe-peak elements
could have originated from pollution by the supernova ejecta.
We find a surprisingly high Li abundance for a star that has evolved
away from the main sequence. We discuss these findings in the context
of companion stars of supernovae.

\end{abstract}

%% Keywords should appear after the \end{abstract} command. The uncommented
%% example has been keyed in ApJ style. See the instructions to authors
%% for the journal to which you are submitting your paper to determine
%% what keyword punctuation is appropriate.


\section{Introduction}

Type Ia supernovae (SNe~Ia) are the best known cosmological distance
indicators at high redshifts. Their use led to the discovery of the
currently accelerating expansion of the Universe (Riess et al. 1998;
Perlmutter et al. 1999); see Filippenko (2005) for a review. They were
also used to reveal an early era of deceleration, up through about 9
billion years after the big bang (Riess et al. 2004, 2007). Larger,
higher-quality samples of SNe~Ia, together with other data, are now
providing increasingly accurate and precise measurements of the dark
energy equation-of-state parameter, $w = P/(\rho c2)$ (e.g., Astier
et al. 2006; Riess et al. 2007; Wood-Vasey et al. 2007; Kowalski
et al. 2008).

Though the increase in the empirical knowledge of SNe~Ia has led to an
enormous advance in their cosmological use, the understanding of the
explosion mechanism still requires careful evaluation (e.g.,
Hillebrandt \& Niemeyer 2000; Wheeler 2007).  While in Type II SNe we
have the advantage that the explosion leaves a compact star to which
surviving companions (if they exist) often remain bound, thus
enabling a large number of 
studies (e.g., Mart{\'\i}n et al. 1992; Israelian et al. 1999), in
SNe~Ia the explosion almost certainly does not produce any compact
object.

To investigate how the explosion takes place, we examine the rates of
SNe~Ia at high redshifts, or we can take a more direct approach and
survey the field of historical SNe Ia. The latter strategy has been
followed since 1997 in a collaboration that used the observatories at
La Palma, Lick, and Keck (Ruiz-Lapuente et al. 2004). The stars
appearing within the 15$\%$ innermost area of the remnant of SN 1572
($0.65'$) were observed both photometrically and spectroscopically at
multiple epochs over seven years. The proper motions of these stars
were measured as well using images obtained with the WFPC2 on board
the {\it Hubble Space Telescope (HST)} (GO--9729). The results
revealed that many properties of any surviving companion star of
SN 1572 were unlike those predicted by hydrodynamical models. For example, 
the surviving companion star (if present) could not be an overluminous 
object nor a blue star, as there were none in that field. Red giants and 
He stars were also discarded as possible companions.

A subgiant star (G2~IV) with metallicity close to solar and moving at high
speed for its distance was suggested as the likely surviving companion 
of the exploding white dwarf (WD) that produced SN 1572 (Ruiz-Lapuente et
al. 2004). This star, denoted Tycho G, has coordinates $\alpha =
00^{\rm h}25^{\rm m}23.7^{\rm s}$ and $\delta = +64^\circ08'02''$
(J2000.0).  Comparisons with a Galactic model showed that the
probability of finding a rapidly moving subgiant with solar
metallicity at a location compatible with the distance of SN 1572 was
very low. A viable scenario for the Tycho SN 1572 progenitor
would be a system resembling the recurrent nova U~Scorpii. This system
contains a  white dwarf of $M_{\rm WD} = 1.55 \pm 0.24$~\Msun and a
secondary star with $M_2 = 0.88 \pm 0.17$~\Msun orbiting with a period 
$P_{\rm orb} \approx 1.23$~d (Thoroughgood et al. 2001).

On the other hand, based on analysis of low-resolution spectra, Ihara
et al. (2007) recently claimed that the spectral type of Tycho G is
not G2~IV as found by Ruiz-Lapuente et al. (2004), but rather
F8~V. In addition, Ihara et al. (2007) suggested the star Tycho~E
as the companion star of the Tycho SN 1572, but we consider their
conclusion to be unjustified. They base it on a single
\ion{Fe}{1} feature at 3720~{\AA} in a spectrum having a signal-to-noise
ratio (SNR) of only $\sim 13$. The short spectral range of their data 
(3600--4200~{\AA}) and the low resolution ($\lambda/\Delta\lambda \approx
400$, with a dispersion of 5~{\AA}/pixel) add to our concern.

The present work resolves questions regarding the
metallicity and spectral classification of Tycho G.
We concentrate on providing both the chemical composition of the
star and the stellar parameters as derived from new data. 
In addition, we present a detailed chemical analysis of Tycho G aimed 
at examining the possible pollution of the companion star by
the supernova. Although more data are still required to definitively
settle this issue, the overall properties of this star remain
consistent with being the surviving companion of SN 1572.

\section{Observations}

In order to perform an abundance analysis of Tycho G, spectra of it were
obtained with the High Resolution Echelle Spectrometer (HIRES; Vogt et
al. 1994) on the Keck~I 10-m telescope (Mauna Kea, Hawaii).  Nine spectra (with
individual exposure times of 1200~s and 1800~s) were obtained on 10 September
2006 (UT dates are used throughout this paper) and four spectra (with exposure
times of 1800~s) on 11 October 2006, covering the spectral regions
3930--5330~{\AA}, 5380--6920~{\AA}, and 6980--8560~{\AA} at resolving power
$\lambda/\Delta\lambda \approx 50,000$. 

The spectra were reduced in a standard
manner using the {\scshape makee} package. We checked in each individual
spectrum the accuracy of the wavelength calibration using the [\ion{O}{1}]
$\lambda$6300.3 and $\lambda$5577.4 night-sky lines and found it to be within
0.2 \kmso. After putting each individual spectrum into the heliocentric frame,
we combined all of the spectra from each night. Then, each night's final
spectrum was cross-correlated with the solar spectrum (Kurucz et al.\ 1984)
properly broadened with the instrumental resolution of $\sim 6.6$~\kmso. The
final HIRES spectrum of Tycho G has a SNR of $\sim 20$ at 5200~{\AA},
$\sim 30$ at 6500~{\AA}, and $\sim 50$ at 7800~{\AA}. This spectrum
was not corrected from telluric lines since it was unnecessary for the
chemical analysis (see \S~\ref{seciron}).

\begin{figure*}[!ht]
%\epsscale{0.70}
%\plotone{f1bw.eps}
\centering
\includegraphics[width=8.cm,angle=0]{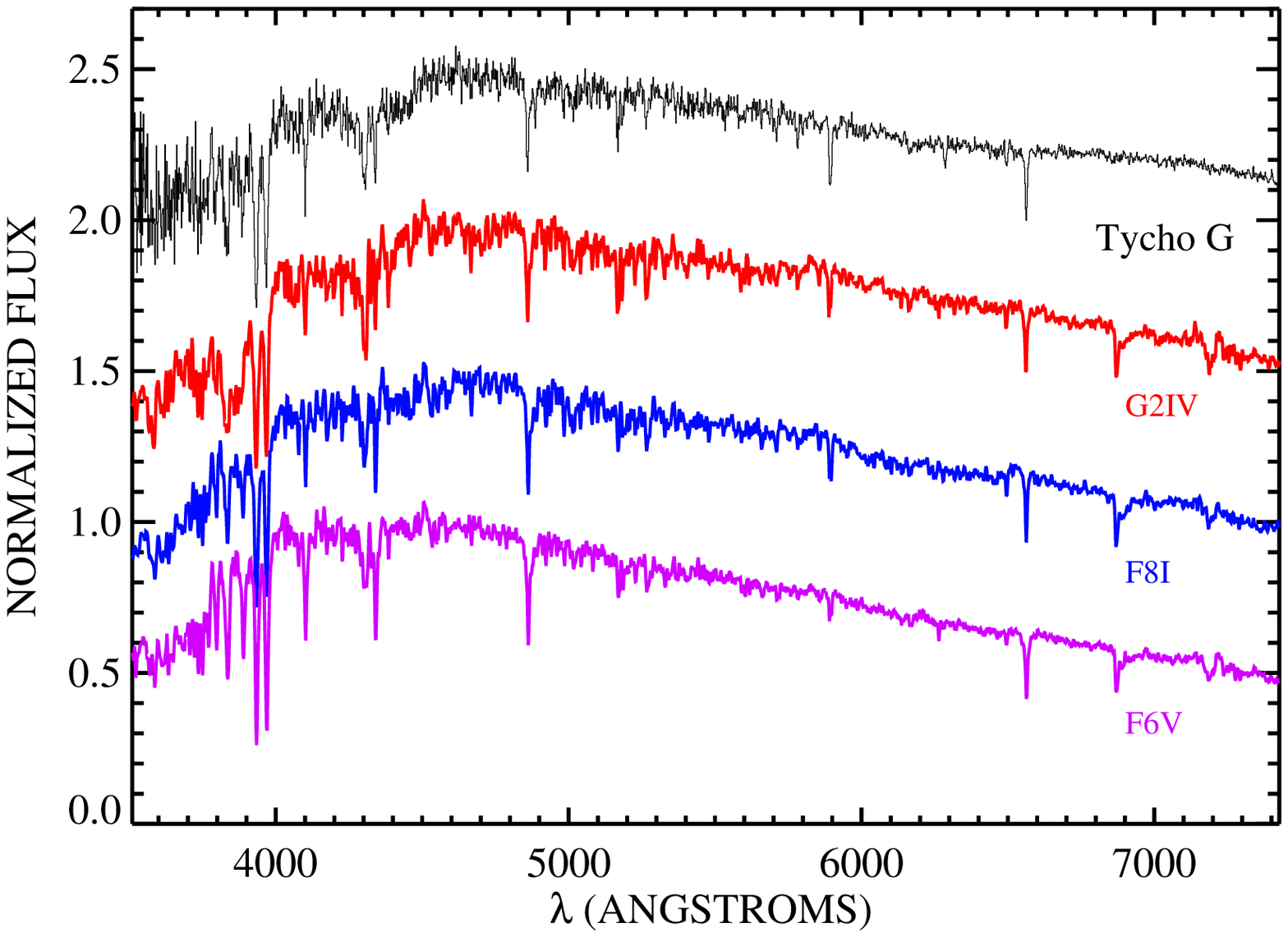}
\includegraphics[width=8.cm,angle=0]{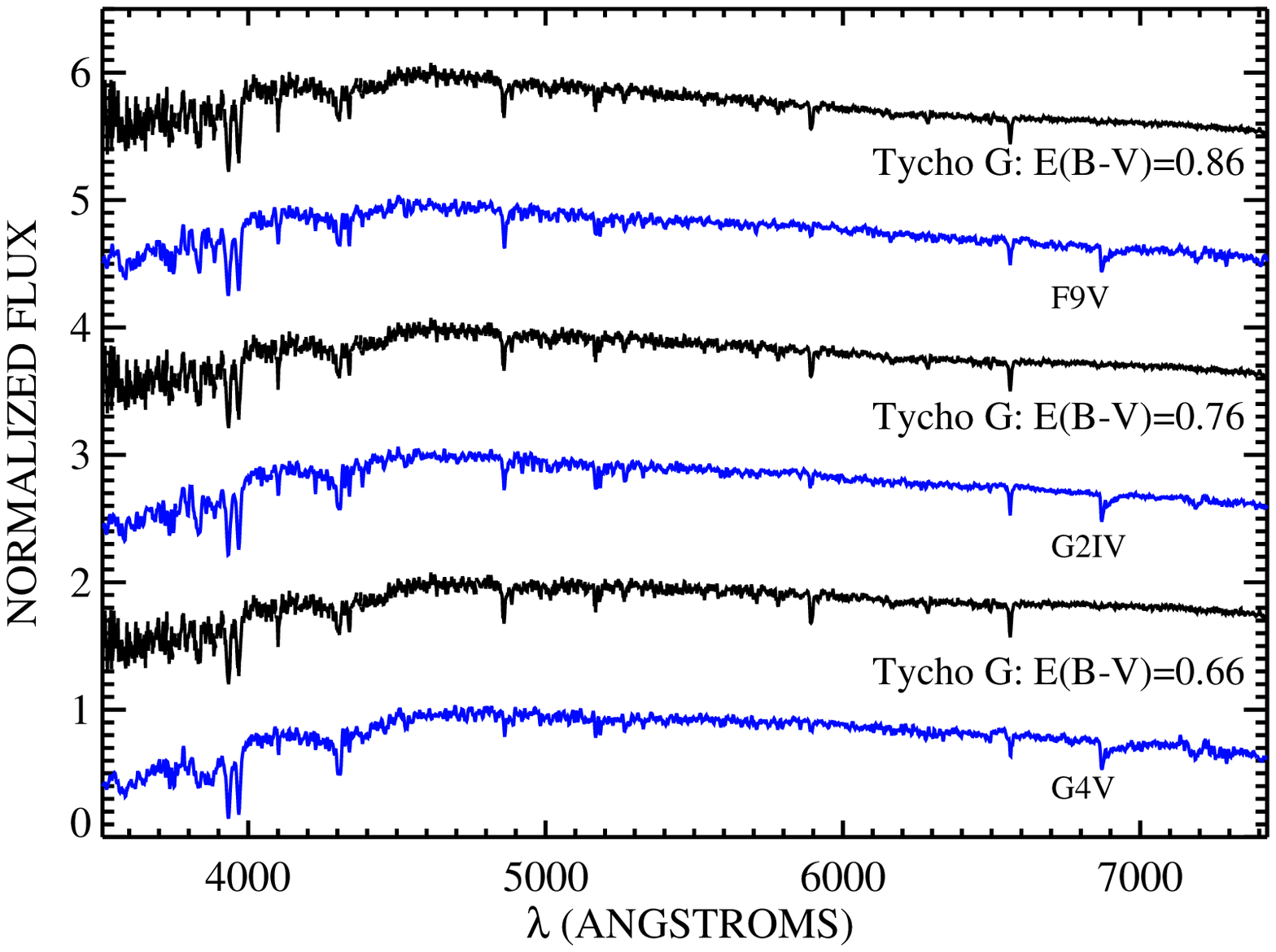}
\caption{Left panel: Keck LRIS spectrum of Tycho G compared with three
template spectra having spectral types G2~IV, F8~I, and F6~V taken
from a library of low-resolution stellar spectra (Jacoby et al.
1984). The absorption feature near 6900~\AA\ in the template
spectra is the telluric B band. Another, weaker telluric feature is
visible near 7200~\AA. Right panel: LRIS spectrum of Tycho G dereddened
with $E(B-V) = 0.66$, 0.76, and 0.86 mag, compared with spectra from
the low-resolution library of Jacoby et al. (1984). The best fits to
the different dereddened spectra of Tycho G are also shown.}
\label{lrtg}
\end{figure*}

With the aim of determining the spectral classification of several
targets in the Tycho SN 1572 field, on 12 December 2007 we obtained
spectra with the Low Resolution
Imaging Spectrometer (LRIS; Oke et 
al. 1995) on the Keck~I telescope. We used the 400/3400 grism, the
400/8500 grating, and the D560 dichroic, covering the range
3180--9150~{\AA}. The full width at half-maximum (FWHM) resolution was
6.4~{\AA} in the red part ($\lambda \gtrsim 5500$~\AA) and 6.3~{\AA} in
the blue, with respective dispersions of 1.86~{\AA} pixel$^{-1}$ and
1.09~{\AA} pixel$^{-1}$. Tycho E, F, G, and D were observed in single
exposures of 700, 200, 400, and 450~s, respectively; see
Ruiz-Lapuente et al. (2004) for star identifications. The SNR of the
individual spectra is 21, 37, 31, and 25 at 4600~{\AA} for Tycho~E, F, G, and D,
respectively. These spectra were reduced using standard techniques (e.g., Foley
et al. 2003), including removal of telluric absorption lines. 

The spectrum of Tycho~G is shown in Figure 1; spectra and
classifications of the other stars are available in
Appendix~\ref{ap1}. In general, our results do not support the
spectral classifications suggested by Ihara et al. (2007), which are
based on low-resolution ($\lambda/\Delta\lambda \approx 400$), low-SNR
spectra obtained with the instrument FOCAS at the Subaru Telescope
over a short spectral range (3600--4200~{\AA}). Moreover, their
relative flux calibrations are questionable since they apply a
correction to the slope of each spectrum that is fitted when comparing
with a template. Thus, we consider their spectral classifications to
be unreliable.

\section{Stellar Parameters}

\subsection{Low-Resolution Spectra}

The blue and red parts of the LRIS spectra were merged and then
rebinned at a scale of 2~{\AA} pixel$^{-1}$. We compared the spectra
of the four targets with a library of low-resolution stellar spectra
from Jacoby, Hunter, \& Christian (1984). We chose this library
because the spectra have a similar resolution ($\sim$4.5~{\AA}), with
a dispersion of 1.4~{\AA} pixel$^{-1}$, and cover a similar spectral
region (3510--7427~{\AA}) as our LRIS observations.

The low-resolution comparison with template spectra depends on the
color excess applied to the LRIS spectra. We dereddened the calibrated
spectra according to the parameterization of Cardelli, Clayton, \&
Mathis (1989), including the update for the near-ultraviolet given by
O'Donnell (1994).  Moreover, there is uncertainty in the spectral
classification of each template.  Thus, the spectral classification of
stars using low-resolution spectra must always be considered only as
an initial guess. The best fit is found for a reddening $E(B-V) =
0.76$ mag, with $\chi2_\nu = 3.68$ (see below for details). This
value of reddening was used in the comparison shown in
Figure~\ref{lrtg}. In the library of Jacoby et al. (1984), there are
not many spectra with luminosity class IV (only F0, F3, G2, G5~IV),
whereas luminosity class V is very well sampled with several templates
for each spectral type. If only low-resolution spectra are used to
derive the spectral and luminosity classes, the results may be
erroneous, since they depend on reddening and on the flux
calibration of the observed spectra.

\begin{figure*}[!ht]
%\epsscale{0.70}
%\plotone{f1bw.eps}
\centering
\includegraphics[width=14.cm,angle=0]{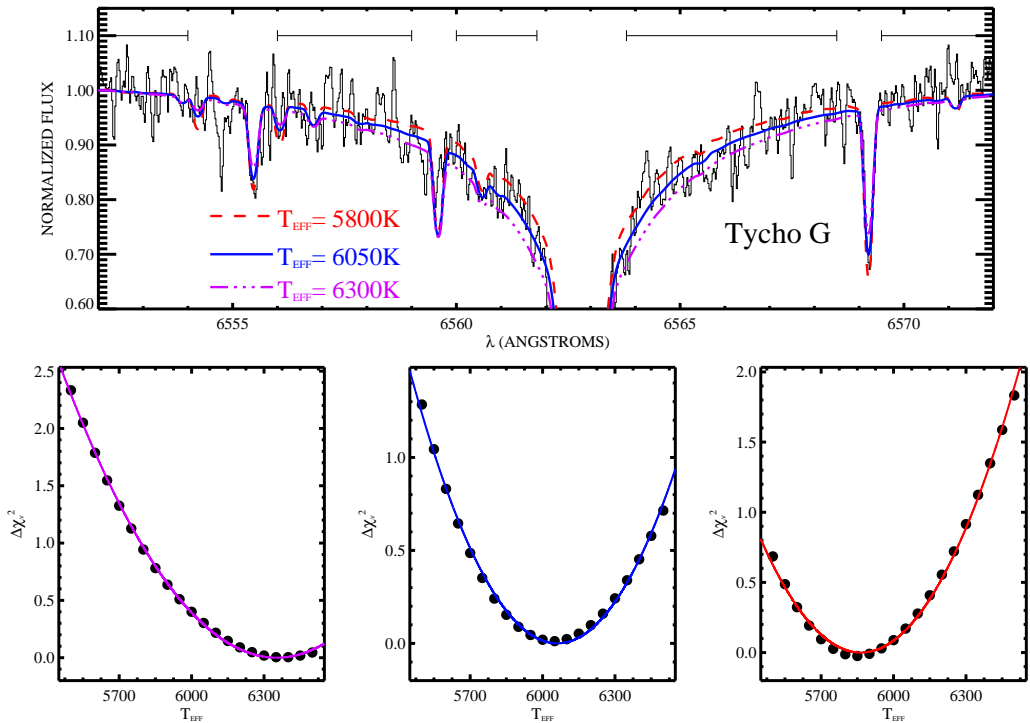}
\caption{Upper panel: Synthetic H$\alpha$ profiles for three effective
temperatures compared to the observed spectrum of \mbox{Tycho G} (SNR
$\approx 30$) and normalized to the level of the observed spectrum of
\mbox{Tycho G} at 6540~{\AA}. We also display the fitting regions
at the top. Lower panel: Results of $\chi2_\nu$
analysis for three continuum positions separated by $\pm 1/{\rm SNR} 
= 0.03$: 1.03 (left), 1.00 (middle), and 0.97 (right).}
\label{halpha}
\end{figure*}

Nonetheless, as a first step, we compare the LRIS spectrum of Tycho G
with the low-resolution spectra of three template stars having
different spectral types (Fig.~\ref{lrtg}). To evaluate the
goodness of fit, we employ a reduced $\chi2$ statistic,
\begin{equation}
\chi2_\nu = \frac{1}{N-M}\sum^N_{i=1}\left(\frac{f_i-F_i}{\sigma_i}\right)2,
\end{equation}
where $N$ is the number of wavelength points, $M$ is the number of free
parameters (here one), $f_i$ is the template flux, $F_i$ is
the target flux (in this case, Tycho G), and $\sigma_i=1/{\rm
SNR}$. The SNR was estimated as a constant
average value in the continuum over the range 4610--4630~{\AA}; this
region was also used to normalize both target and template
spectra. The spectral regions used to evaluate $\chi2$ are
3850--6800~{\AA} and 7000--7400~{\AA}, avoiding strong telluric lines
near 6900~{\AA} present only in the template spectra.

The best fit is found for the G2~IV template with $\chi2_\nu = 3.68$,
whereas the F8~I and F6~V templates provide $\chi2_\nu = 6.39$ and
9.66, respectively. Ihara et al. (2007) have claimed that the spectral
type of Tycho G could be either F8~I, F8~V, or F6~V, rather than G2~IV
as found by Ruiz-Lapuente et al. (2004). However, our analysis shows
that those other spectral types provide worse fits than the G2~IV
template.

\begin{figure*}[!ht]
%\epsscale{0.70}
%\plotone{f1bw.eps}
\centering
\includegraphics[width=11.cm,angle=90]{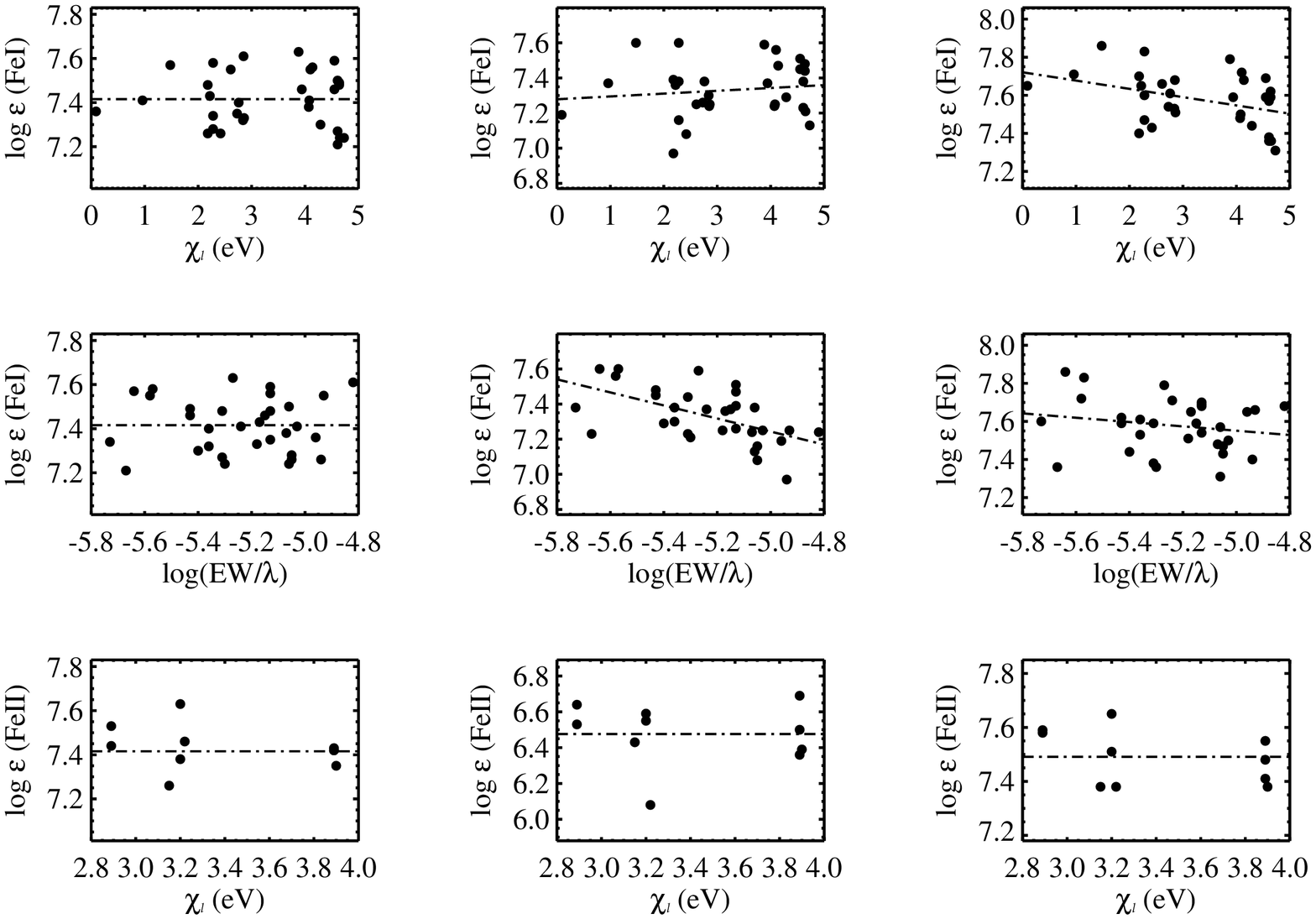}
\caption{\ion{Fe}{1} and \ion{Fe}{2} abundances computed using the
best-fit atmospheric parameters of \mbox{Tycho G} (left panels) and two sets of
stellar parameters: \teff$=5900$ K and \loggl\ $=2$ dex (middle panels),
\teff $=6200$ K and \loggl\ $=4.35$ dex (right panels). Upper panels:
\ion{Fe}{1} abundance vs. excitation potential; middle panels:
\ion{Fe}{1} abundance vs. reduced equivalent widths; bottom panels:
\ion{Fe}{2} abundance vs. excitation potential. The dashed lines in
the upper and middle panels represent the fits to the points, and in
the bottom panels the mean abundance.}    
\label{fig1}
\end{figure*}

In the following sections we use more accurate classification methods,
which are independent of reddening, to derive not only the spectral
type of Tycho G but actually the effective temperature and surface
gravity, using a high-resolution spectrum.

\subsection{H$\alpha$ Profile}

The wings of H$\alpha$ are a very good temperature indicator 
(e.g., Barklem et al. 2002). Adopting the theory of Ali \& Griem (1965,
1966) for resonance broadening and Griem (1960) for Stark broadening,
we computed H$\alpha$ profiles for several effective temperatures,
using the code SYNTHE (Kurucz et al. 2005; Sbordone et al. 2005). For
further details on the computations of hydrogen lines in SYNTHE,
see Castelli \& Kurucz (2001) and Cowley \& Castelli (2002).
 
Figure~\ref{halpha} compares these synthetic H$\alpha$ profiles with
the observed profile for several temperatures. To evaluate the
goodness of fit, we employ the same method as in the previous section.
However, in this case $M = 1$ (the effective temperature \teffo),
$f_i$ is the synthetic normalized flux, $F_i$ is the observed 
normalized flux, and $\sigma_i=1/{\rm SNR}$. The SNR was estimated as
a constant average value in continuum regions close to the observed H$\alpha$
profile. The fitting regions are indicated in Figure~\ref{halpha},
which contains all of the spectral regions close the center of the
H$\alpha$ profile where there are no stellar lines and where the
normalized flux is greater than $\sim0.7$. The best fit provided a
temperature of \teff $=6050\pm250$ K, with the error bars estimated
from the best fits after displacing the observed continuum up and down
by $1/{\rm SNR}=0.03$. 

In the following section, we adopt the currently most reliable method
for the determination of the stellar parameters.

\subsection{Ionization Equilibrium of Iron\label{seciron}} 

The best determination of the atmospheric parameters of \mbox{Tycho G}
can be obtained from the ionization equilibrium of iron. We measured the 
equivalent widths of 32 \ion{Fe}{1} and 10 \ion{Fe}{2} isolated lines using 
an automatic line-fitting procedure which performs both line detection
and Gaussian fits to unblended lines (Fran\c cois et al. 2003). The
analysis was done using the 2002 version of the code MOOG (Sneden
1973) and a grid of local thermodynamic equilibrium (LTE) model
atmospheres (Kurucz 1993). We adopted the atomic data from Santos,
Israelian, \& Mayor (2004), where the $\log gf$ values are adjusted 
until the solar atlas (Kurucz et al.\ 1984) is reproduced with a Kurucz 
model for the Sun having \teff$=5777$ K, \loggl\ $=4.44$ dex, $\xi_t=1.00$ \kmso,
and $\log_\odot \epsilon(Fe)=7.47$ dex. This line list was designed
to determine stellar parameters of planet-host stars with roughly the
same atmospheric parameters and metallicity as Tycho G; it
provides a compilation of 42 completely unblended Fe lines ideal
for the determination of stellar atmospheric parameters. Note that
none of the stellar lines used in this work for chemical analysis is 
affected by telluric lines.

The atmospheric parameters were
determined by iterating until correlation coefficients between $\log
\epsilon$(\ion{Fe}{1}) and $\chi_l$, as well as between $\log
\epsilon$(\ion{Fe}{1}) and $\log (W_\lambda/\lambda)$, were zero, and
the mean abundances from \ion{Fe}{1} and \ion{Fe}{2} lines were the
same (see left panels of Fig.~\ref{fig1}). The derived parameters were
$T_{\mathrm{eff}} = 5900 \pm 100$ K, \loggl$= 3.85 \pm
0.30$ dex, $\xi_t=1.23\pm0.23$ \kmso, and $\mathrm{[Fe/H]} = -0.05 \pm
0.09$. The uncertainties in the stellar parameters were estimated as
described by Gonzalez \& Vanture (1998). Thus, the uncertainties in $\xi_t$ and
\teff take into account the standard deviation of the slope of the least-squares
fits, $\log \epsilon$(\ion{Fe}{1}) versus $\log (W_\lambda/\lambda)$ and $\log
\epsilon$(\ion{Fe}{1}) versus $\chi_l$. The uncertainty in \logg\
considers the uncertainty on \teff in addition to the scatter of the \ion{Fe}{2}
abundances. Finally, the uncertainty in Fe abundance is estimated from a
combination of the uncertainties in $\xi_t$ and \teff, in addition to the scatter
of the \ion{Fe}{1} abundances, all added in quadrature.

\begin{figure*}[!ht]
%\epsscale{0.70}
%\plotone{f1bw.eps}
\centering
\includegraphics[width=16cm,angle=0]{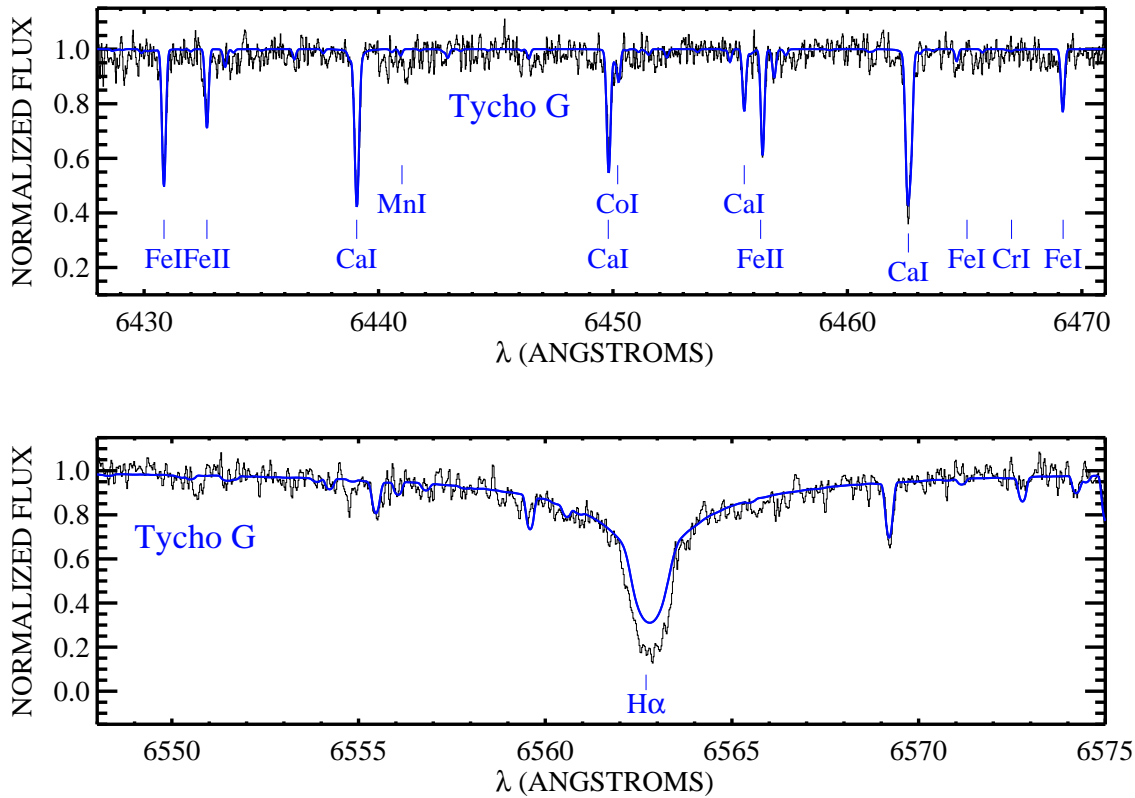}
\caption{Best synthetic spectral fits (solid lines) to the Keck 
HIRES spectrum of \mbox{Tycho G} for two spectral regions. The
core of the H$\alpha$ line suffers from NLTE effects that are not 
taken into account in our LTE synthetic spectrum.}   
\label{fig2}
\end{figure*}

In Figure \ref{fig1}, we also show two set of stellar parameters,
\teff $=5900$ K and \loggl\ $=2$ dex (middle panels), and \teff $=6200$ K
and \loggl\ $=4.35$ dex (right panels), corresponding to the spectral
types F8~I and F6-7~V (Gray 1992). The lower-middle and upper-middle
panels show inconsistent abundances between \ion{Fe}{1} and
\ion{Fe}{2}, whereas the upper-right panel shows a negative slope for
iron abundance versus excitation potential $\chi_l$ of the spectral
lines \ion{Fe}{1}. Therefore, these two sets of parameters cannot be
the stellar parameters of Tycho G. According to the spectral
classification provided by Gray (1992), the stellar parameters of
Tycho G correspond to a spectral type of G0-1~IV.
 
Non-LTE (NLTE) corrections for stars with similar stellar parameters
and metallicity have been estimated at $\la 0.07$ dex for \ion{Fe}{1}
lines, while \ion{Fe}{2} seems to be insensitive to NLTE effects
(Th\'evenin \& Idiart 1999). In addition, Santos et al. (2004)
compared spectroscopic surface gravities with $\log g$ using Hipparcos
parallaxes and found an average difference of $\sim+0.03$ dex, which
might in fact reflect NLTE effects on \ion{Fe}{1} lines. Our
assumption of LTE seems to be accurate enough to determine the surface
gravity of the star within the error bars. Note that ionization 
equilibrium also holds for Si and Cr (see \S~\ref{secabun}). 

This determination of effective temperature is consistent within the
error bars with that derived from the H$\alpha$ profile (\teff
$=6050\pm250$ K). However, echelle spectrographs are not ideally
suited for measuring the shape of broad features such as the wings of
Balmer lines (Allende Prieto et al. 2004), and the SNR of $\sim$30 is
not high enough for the determination of effective temperature from the
H$\alpha$ profile with great accuracy. For this reason, we will adopt
the \teff $=5900 \pm 100$ K value for the determination of the chemical
abundances. 

\begin{figure*}[!ht]
\centering
\includegraphics[width=16cm,angle=0]{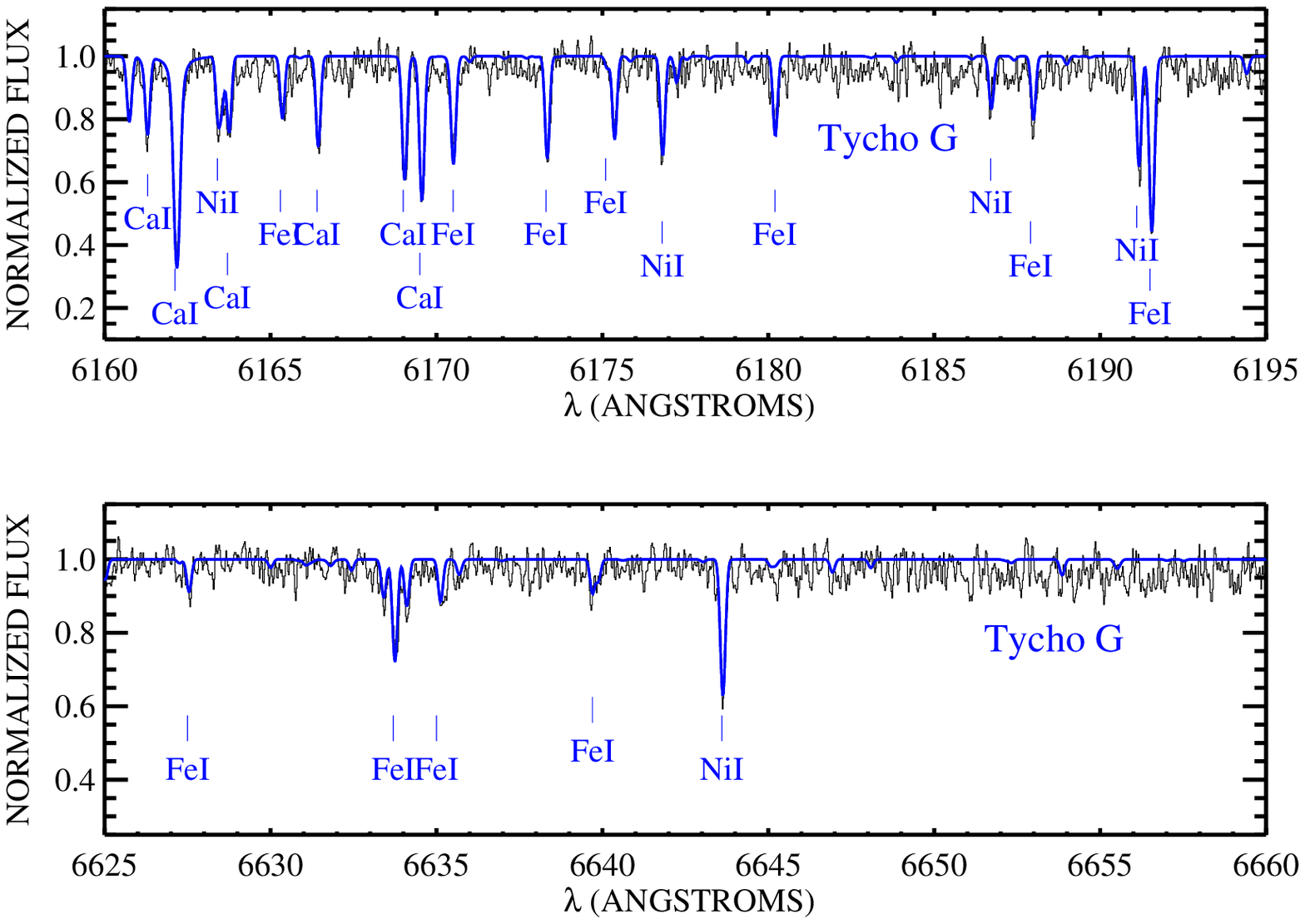}
\caption{Same as Fig.~\ref{fig2}, but for other spectral
regions.}   
\label{fig3b}
\end{figure*}

\begin{figure*}[!ht]
\centering
\includegraphics[width=16cm,angle=0]{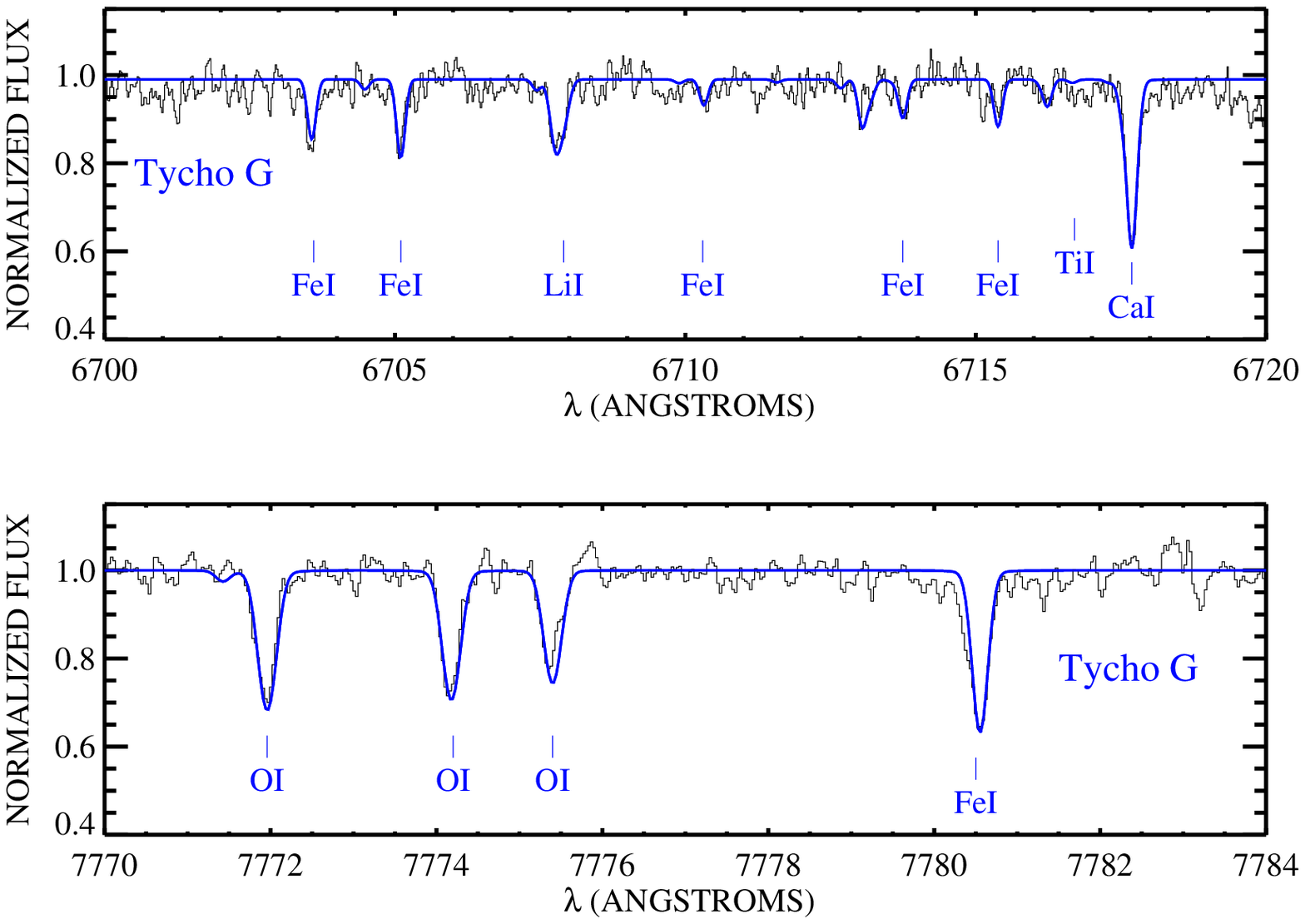}
\caption{Same as Fig.~\ref{fig2}, but for other spectral
regions.}   
\label{fig3}
\end{figure*}

\section{Chemical Abundances\label{secabun}}

Using the derived stellar parameters, we determine the element
abundances from the absorption-line equivalent width (EW) measurements
of the elements given in Table~\ref{tblB}, except for lithium and
sulfur for which we use a $\chi2$ procedure to find the best fit to
the observed features (e.g., Gonz\'alez Hern\'andez et al. 2004, 2005,
2006). The atomic data for O, C, S, and Zn are taken from Ecuvillon,
Israelian, \& Santos (2004, 2006), 
those of Na, Mg, Al, Si, Ca, Sc, Ti, V, Cr, Mn, 
Co, and Ni are from Gilli, Israelian, \& Ecuvillon (2006), and 
those for Sr, Y, and Ba are from Reddy et al. (2003).

In Figures~\ref{fig2},~\ref{fig3b}, and~\ref{fig3} we display several 
spectral regions
of the observed spectrum of \mbox{Tycho G} in comparison with
synthetic spectra computed using the derived element
abundances. Uncertainties in the abundances of all elements were then
determined, adding in quadrature the errors due to the sensitivities
of the resulting abundances to changes in assumed atmospheric
parameters and the dispersion of the abundances from individual lines
of each element.

In Table~\ref{tbl2} we provide the average abundance of each element
together with the errors. The errors in the element abundances show
their sensitivity to the uncertainties in the effective temperature
($\Delta_{T_{\mathrm{eff}}}$), surface gravity ($\Delta_{\log g}$),
microturbulence ($\Delta_{\xi}$), and the dispersion of the
measurements from different spectral features ($\Delta_{\sigma}$). The
errors $\Delta_{\sigma}$ were estimated as $\Delta_{\sigma}
=\sigma/\sqrt{N}$, where $\sigma$ is the standard deviation of the $N$
measurements.

\begin{deluxetable}{lrrrrrrrrrrr}
\tabletypesize{\scriptsize}
\tablecaption{Chemical Abundances of Tycho G}
\tablewidth{0pt}
\tablehead{Species & $\log \epsilon(\mathrm{X})_{\odot}$\tablenotemark{a} & 
$[{\rm X}/{\rm H}]$ & $[{\rm X}/{\rm Fe}]$ & 
${\sigma}$ & $\Delta_{\sigma}$ &  
$\Delta_{T_{\rm eff}}$ & $\Delta_{\log g}$ &
$\Delta_{\xi}$ & $\Delta \mathrm{[X/H]}$ & $\Delta \mathrm{[X/Fe]}$ &
$n$\tablenotemark{b}} 
\startdata
C~I           & 8.56 & -0.15 & -0.07  &  0.15 &  0.07 & -0.06 & 0.09  &   0   & 0.13 & 0.19 & 5 \\ 
O~I           & 8.74 & -0.02 &  0.03  &  0.02 &  0.01 & -0.08 & 0.07  & -0.02 & 0.11 & 0.18 & 3 \\ 
Na~I          & 6.33 &  0.04 &  0.09  &  0.18 &  0.10 &  0.05 & -0.05 & -0.02 & 0.12 & 0.12 & 3 \\ 
Mg~I          & 7.58 &  0.08 &  0.13  &  0.06 &  0.03 &  0.05 & -0.04 & -0.03 & 0.08 & 0.06 & 4 \\ 
Al~I          & 6.47 &  0.16 &  0.21  &  0.05 &  0.03 &  0.06 & -0.01 & -0.01 & 0.07 & 0.05 & 2 \\ 
Si~I          & 7.55 & -0.03 &  0.02  &  0.09 &  0.02 &  0.04 & -0.01 & -0.02 & 0.05 & 0.05 & 12 \\ 
Si~II         & 7.55 & -0.03 &  0.02  &  --   &  --   & -0.08 &  0.11 & -0.03 & 0.14 & 0.20 & 6371 \\ 
S~I\tablenotemark{c} & 7.21 & -0.02 &  0.03  &  0.07 &  0.05 & -0.03 &  0.12 &     0 & 0.13 & 0.18 & 2 \\ 
Ca~I          & 6.36 & -0.04 &  0.01  &  0.11 &  0.03 &  0.08 & -0.04 & -0.05 & 0.11 & 0.05 & 10 \\ 
Sc~II         & 3.10 & -0.06 & -0.01  &  0.09 &  0.03 &  0.01 &  0.12 & -0.04 & 0.13 & 0.15 & 7 \\ 
Ti~I          & 4.99 & -0.03 &  0.02  &  0.14 &  0.04 &  0.11 &     0 & -0.02 & 0.12 & 0.06 & 10 \\ 
V~I           & 4.00 &  0.07 &  0.12  &  0.21 &  0.07 &  0.10 & -0.01 & -0.01 & 0.12 & 0.08 & 8 \\ 
Cr~I          & 5.67 & -0.04 &  0.01  &  0.16 &  0.05 &  0.07 &     0 & -0.01 & 0.09 & 0.06 & 9 \\ 
Cr~II         & 5.67 & -0.03 &  0.02  &   --  &   --  & -0.02 &  0.11 & -0.04 & 0.12 & 0.16 & 5305 \\ 
Mn~I          & 5.39 &  0.02 &  0.07  &  0.15 &  0.05 &  0.09 & -0.02 & -0.05 & 0.12 & 0.06 & 8 \\ 
Fe~I          & 7.47 & -0.05 &  0     &  0.12 &  0.02 &  0.08 & -0.01 & -0.04 & 0.09 &  --  & 32 \\ 
Fe~II         & 7.47 & -0.05 &  0     &  0.10 &  0.03 &  0    &  0.13 & -0.05 & 0.14 & 0.17 & 10 \\ 
Co~I          & 4.92 &  0.13 &  0.18  &  0.14 &  0.07 &  0.09 &    0  & -0.02 & 0.12 & 0.08 & 4 \\ 
Ni~I          & 6.25 &  0.11 &  0.16  &  0.19 &  0.04 &  0.08 & -0.01 & -0.04 & 0.10 & 0.04 & 22 \\ 
Zn~I          & 4.60 & -0.06 & -0.01  &  0.09 &  0.05 &  0.03 &  0.02 & -0.08 & 0.10 & 0.09 & 3 \\ 
Sr~I          & 2.64 &  0.07 &  0.12  &  --   &  --   &  0.09 & -0.02 & -0.06 & 0.11 & 0.03 & 4607 \\ 
Y~II          & 2.12 & -0.14 & -0.09  &  0.16 &  0.08 &  0.02 &  0.12 & -0.06 & 0.16 & 0.17 & 4 \\ 
Ba~II         & 2.20 &  0.19 &  0.24  &  0.12 &  0.07 &  0.03 &  0.05 & -0.17 & 0.19 & 0.17 & 3 \\ 
\enddata

\tablecomments{Chemical abundances of Tycho~G and uncertainties produced
by $\Delta({T_{\rm eff}}) = +100$\,K, $\Delta({\log g}) = +0.3$ dex, and
$\Delta({\xi}) = +0.23~$\kmso.}

\tablenotetext{a}{The solar element abundances were adopted from
Santos et al. (2004), Ecuvillon et al. (2004, 2006), Gilli et al.
(2006), and Reddy et al. (2003).} 

\tablenotetext{b}{Number of spectral lines of this element analyzed in
the star, or if there is only one, its wavelength.}

\tablenotetext{c}{These abundances were determined by fitting the
observed spectra with synthetic spectra computed with the LTE code
MOOG.} 

\label{tbl2}      
\end{deluxetable}

\begin{figure*}[!ht]
\centering
\includegraphics[width=11cm,angle=+90]{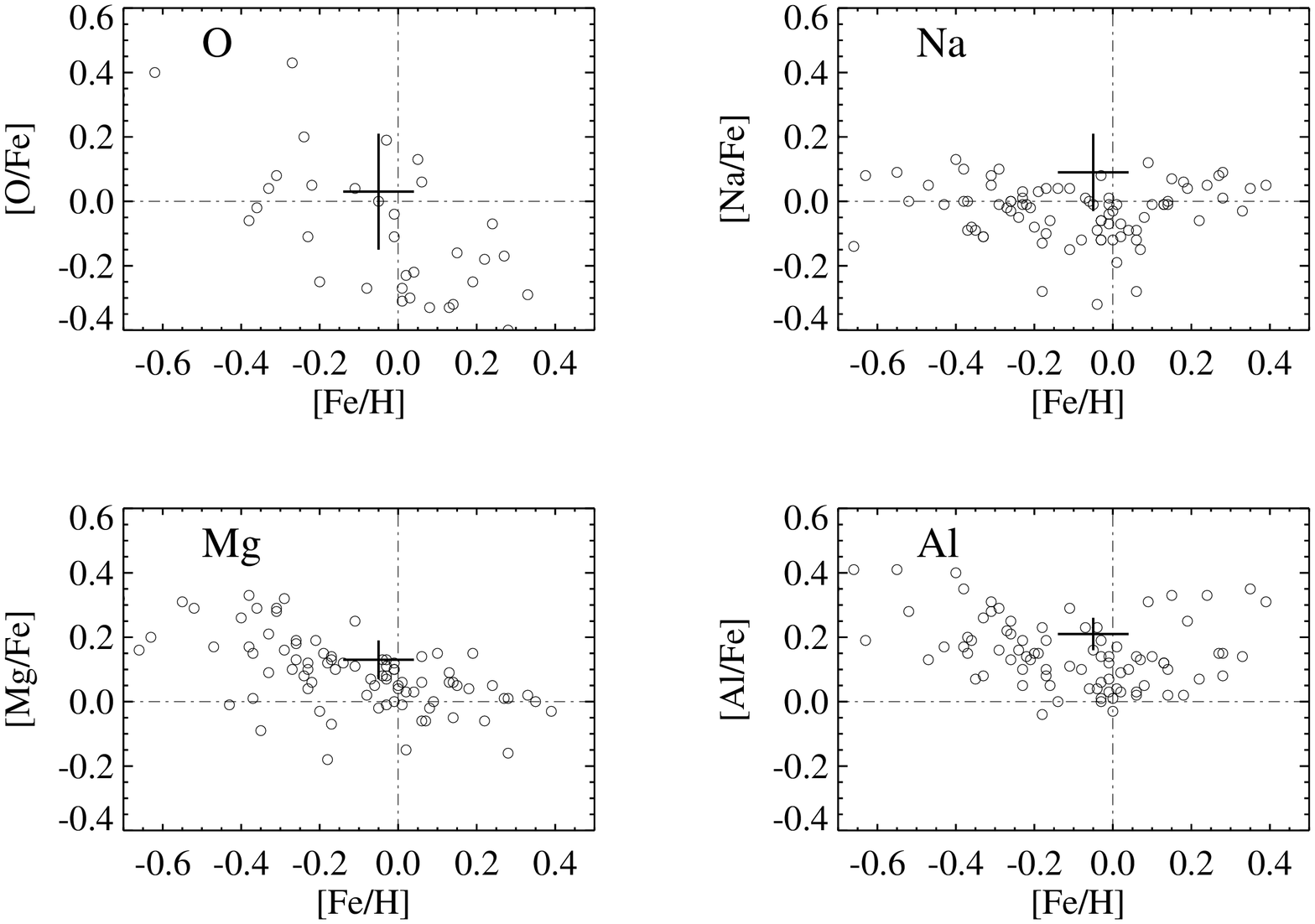}
\caption{Abundance ratios of Tycho G (wide cross) in comparison with
the abundances of G and K metal-rich dwarf stars. Galactic trends were
taken from Ecuvillon et al. (2004), Ecuvillon et al. (2006), and Gilli
et al. (2006). The size of the cross indicates the $1\sigma$
uncertainty. For the abundance of oxygen in
metal-rich dwarfs, we have only considered abundance measurements in
NLTE for the triplet \ion{O}{1} 7771--5 {\AA}. The dashed-dotted lines
indicate solar abundance values.}  
\label{fig4}
\end{figure*}

\begin{figure*}[!ht]
%\epsscale{0.70}
%\plotone{f1bw.eps}
\centering
\includegraphics[width=11cm,angle=90]{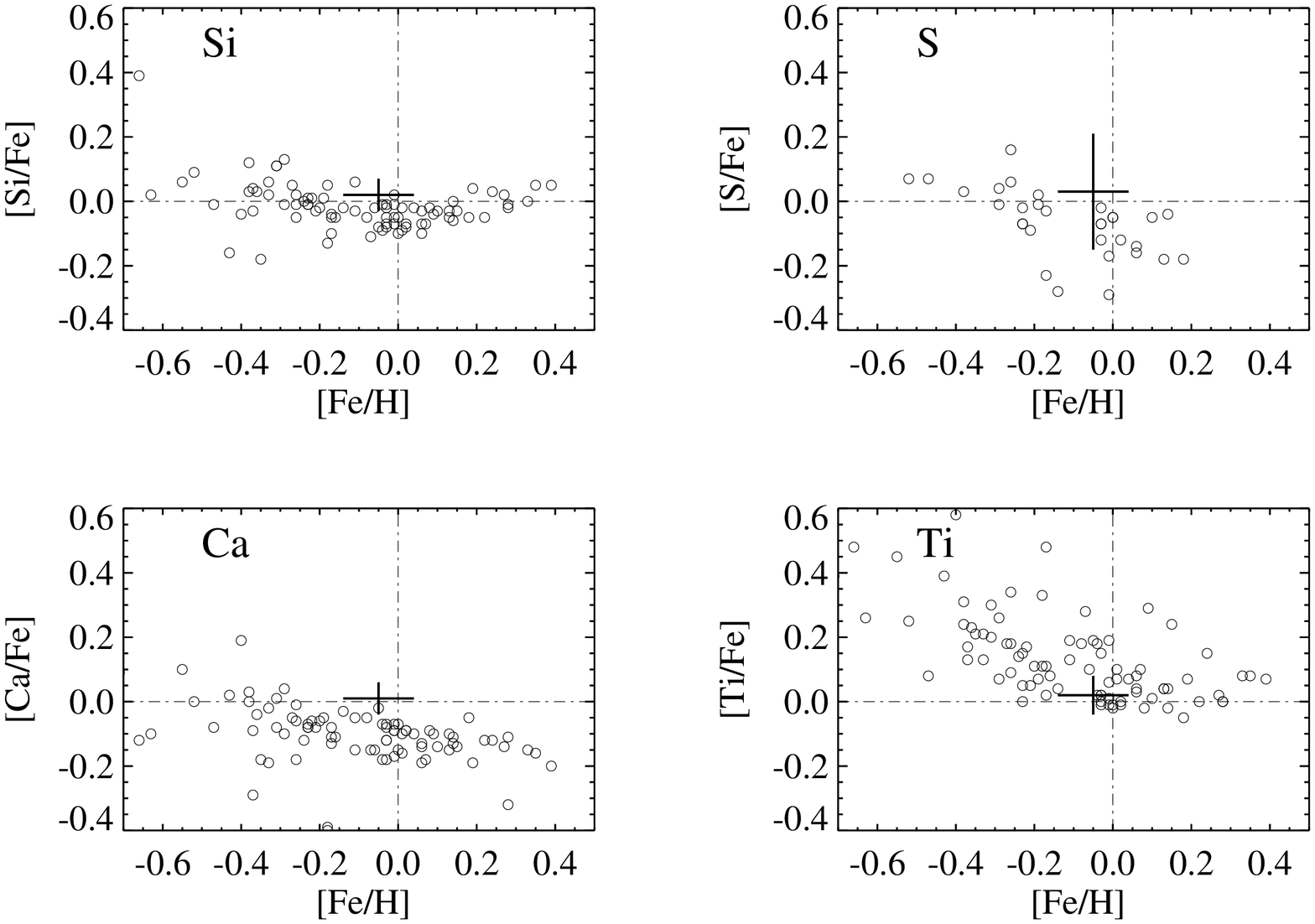}
\caption{Same as Fig.~\ref{fig4}, but for different elements.} 
\label{fig5}
\end{figure*}

\begin{figure*}[!ht]
%\epsscale{0.70}
%\plotone{f1bw.eps}
\centering
\includegraphics[width=11cm,angle=90]{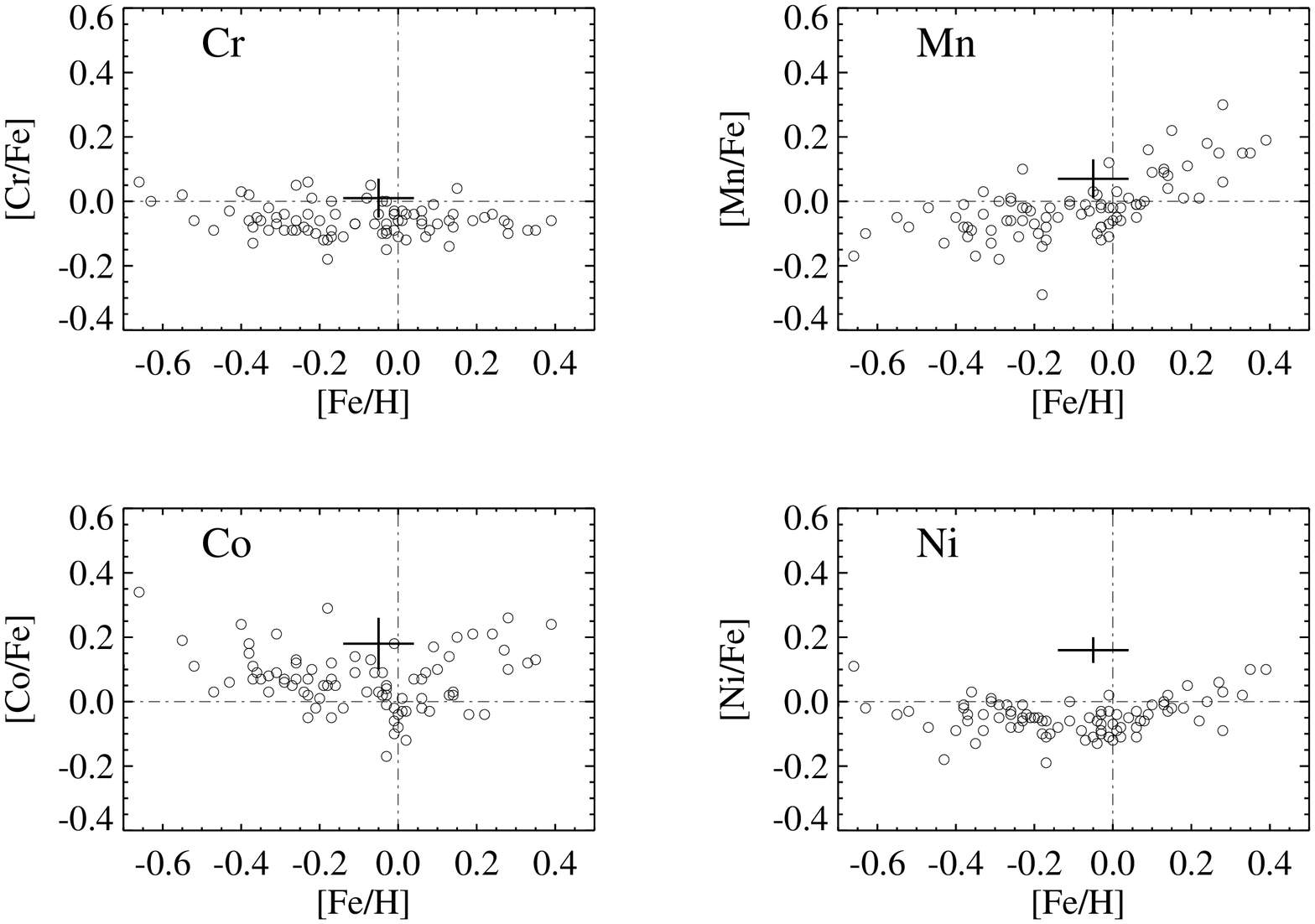}
\caption{Same as Fig.~\ref{fig4}, but for different elements.} 
\label{fig6}
\end{figure*}

The abundances of several elements listed in Table~\ref{tbl2} relative
to iron are compared in Figures~\ref{fig4},~\ref{fig5}, and~\ref{fig6}
with the Galactic trends of these elements in the relevant range of
metallicities. The abundances of some heavy elements relative to iron
are above the solar values. The value found for [Ni/Fe], $0.16 \pm
0.04$, is well above the average value if we consider the intrinsic
dispersion ([Ni/Fe] $= -0.07 \pm 0.04$, see Fig.~\ref{fig6}) for the
stars with similar metallicity of Tycho G ([Fe/H] $= -0.05 \pm 0.09$).
In Figure~\ref{fig3b} we display several spectral ranges where some
\ion{Ni}{1} lines are well reproduced.
Generally, Ni tracks Fe throughout the 
[Fe/H] range down to [Fe/H] $=-1$. The average value and scatter
found by Reddy, Lambert, \& Allende Prieto (2006) is [Ni/Fe] $= -0.05
\pm 0.02$ for thin-disk stars and [Ni/Fe] $= -0.01 \pm 0.04$ for
thick-disk stars. Bensby et al. (2005) found [Ni/Fe] $= -0.06
\pm 0.04$ for thin-disk stars and [Ni/Fe] $= -0.02\pm 0.02$ for
thick-disk stars. This suggests pollution in Ni by Tycho SN 1572
ejecta. We find in 
general solar or slightly above solar abundances of heavy elements,
while the $\alpha$ elements are consistent with the solar value or
below.

The binary companions of black holes or neutron stars show a different
enhancement of metals with respect to the solar values. These stars
exhibit significant enrichment in $\alpha$ elements. For
instance, the black hole binary Nova Scorpii 1994 ([Fe/H] = --0.1) shows
enhancements of [$\alpha$/Fe] = 4--8 in Mg, S, Si, and O (Gonz\'alez
Hern\'andez et al. 2008a). The pollution 
seems to be related to the fallback of the ejected material of the
supernova. In SNe~Ia, one does not expect fallback of material onto the
companion star since the compact object (the white dwarf in this case)
is destroyed and the gravitational potential well is not sufficiently
deep to retain the ejected material. Therefore, one expects to have a
low contamination in intermediate-mass elements. The material captured
by the companion should consist of heavy elements; they are less
likely to escape the companion star since they move at lower
velocities in the supernova ejecta. A discussion of this is given in
\S~\ref{polluted}.

Another feature of interest in Tycho G is its high lithium abundance;
the Li line at 6708~\AA\ is pronounced. Modeling the Li abundance
gives $A$(Li) = $2.50 \pm 0.09$. The lithium abundance is provided
here as $A$(Li) $=\log[N({\rm Li})/N({\rm H})]+12$ and takes into
account NLTE effects. The NLTE abundance correction,
$\Delta_\mathrm{NLTE}= \log \epsilon(\mathrm{X})_\mathrm{NLTE}-\log
\epsilon(\mathrm{X})_\mathrm{LTE}$, for Li was derived from the
theoretical LTE and NLTE curves of growth in Pavlenko \& Magazz\`u
(1996).

The high Li abundance in Tycho G is intriguing since Li is easily
destroyed in the convective envelopes of stars that have evolved away
from the main sequence. Subgiants with $A$(Li) $>$ 2.2 are rare.
Thor\'en, Edvardsson, \& Gustafsson (2004) found that subgiants with
detected Li divide into two groups: the stars with $T_{\rm eff} <
5500$~K have $A$(Li) $\le$ 1.5, and a warmer group of stars have
higher abundances, although still much lower than their primordial
abundances, with $A$(Li) $\le$ 2.2. This seems to go in the direction
of confirming post-main-sequence evolution toward lower amounts of Li
as expected from stellar models. 

Tycho G shares a high Li content with a known sample of binary
companions of neutron stars and black holes (Mart{\'\i}n et al. 1994a;
Gonz\'alez Hern\'andez et al. 2004, 2005).  However, we know that
cataclysmic variables, whose compact objects are white dwarfs, do not
show such high Li abundances (Mart{\'\i}n et al.  1995). We discuss
this further in \S~\ref{licap}.

\begin{deluxetable}{lrrrrrr}
\tabletypesize{\scriptsize}
\tablecaption{Distance of Tycho G\label{tbl3}}
\tablewidth{0pt}
\tablehead{\colhead{Parameter} & \colhead{$E(B-V)\tablenotemark{a}=0.76$}  &
\colhead{$E(B-V)\tablenotemark{a}=0.66$} & \colhead{\teff$\tablenotemark{b}=5800$\,K}  &
\colhead{\logg$\tablenotemark{c}=4.15$}  & \colhead{\logg$\tablenotemark{d}=3.55$}  &
\colhead{$M_{\star}\tablenotemark{e}=0.8~\Msuno$}} 
\startdata
$R_{\star} (\Rsuno)$   &  1.87          & 1.87           & 1.87 	  & 1.32 	   & 2.63 	    & 1.67     \\
$L_{\star}$ (\Lsuno)   &  3.80          & 3.80           & 3.54f 	  & 1.90 	   & 7.57	    & 3.04     \\
$d_{m_V}\tablenotemark{f}$ (kpc)      &  $3.80\pm0.07$ & $4.39\pm0.08$  & $3.65\pm0.07$  & $2.69\pm0.05$  & $5.38\pm0.10$  & $3.40\pm0.06$ \\
$d_{m_R}$ (kpc)          &  $3.52\pm0.05$ & $3.98\pm0.06$  & $3.41\pm0.05$  & $2.49\pm0.03$  & $4.99\pm0.07$  & $3.15\pm0.04$ \\
$d_{m_J}$ (kpc)          &  $3.49\pm0.11$ & $3.65\pm0.12$  & $3.43\pm0.11$  & $2.47\pm0.08$  & $4.94\pm0.16$  & $3.12\pm0.10$ \\
$d_{m_H}$ (kpc)          &  $3.09\pm0.15$ & $3.22\pm0.16$  & $3.06\pm0.15$  & $2.20\pm0.11$  & $4.37\pm0.22$  & $2.77\pm0.14$ \\
$d_{m_K}$ (kpc)          &  $3.16\pm0.18$ & $3.25\pm0.18$  & $3.13\pm0.18$  & $2.24\pm0.13$  & $4.46\pm0.25$  & $2.82\pm0.16$ \\
$d_{\rm av}^{g}$ (kpc) &  $3.50\pm0.31$ & $3.85\pm0.52$  & $3.40\pm0.25$  & $2.48\pm0.21$  & $4.95\pm0.44$  & $3.13\pm0.27$ \\
\enddata
\tablenotetext{a}{The distance is estimated by adopting \teff $=5900$ K, 
\loggl\ $=3.85$ dex, $M_{\star}=1$ \Msuno, and the color excess
indicated in each column by $E(B-V)$.}

\tablenotetext{b}{The distance is estimated by adopting \teff $=5800$ K, 
\loggl\ $=3.85$ dex, $M_{\star}=1$ \Msuno, and $E(B-V)=0.76$ mag.}

\tablenotetext{c}{The distance is estimated by adopting \teff $=5900$ K, 
\loggl\ $=4.15$ dex, $M_{\star}=1$ \Msuno, and $E(B-V)=0.76$ mag.}

\tablenotetext{d}{The distance is estimated by adopting \teff $=5900$ K, 
\loggl\ $=3.55$ dex, $M_{\star}=1$ \Msuno, and $E(B-V)=0.76$ mag.}

\tablenotetext{e}{The distance is estimated by adopting \teff $=5900$ K, 
\loggl\ $=3.85$ dex, $M_{\star}=0.8$ \Msuno, and $E(B-V)=0.76$ mag.}

\tablenotetext{f}{The error bar in the distance is estimated from the
uncertainty in the magnitude.}

\tablenotetext{g}{Average distance weighted by the uncertainty of each individual
distance determination. The error bar shows the large dispersion 
among the individual determinations.}
\end{deluxetable}

\section{Distance\label{dist}}

One can estimate the distance of Tycho G from different photometric
colors and stellar parameters. We use the photometric magnitudes in five
different filters: $m_V=18.71\pm0.04$ and $m_R=17.83\pm0.03$ mag from
Ruiz-Lapuente et al. (2004), and $m_J=15.84\pm0.07$,
$m_H=15.16\pm0.11$, and $m_K=15.03\pm0.12$ mag from the 2MASS
catalog\footnote{The Two Micron All Sky Survey is a joint project of
the University of Massachusetts and the Infrared Processing and
Analysis Center/California Institute of Technology, funded by the
National Aeronautics and Space Administration (NASA) and the National
Science Foundation (NSF).}. We derived the radius of the star from the
surface gravity, \loggl\ $=3.85\pm0.3$ dex, and assuming a mass of 1
\Msuno. This radius, together with the spectroscopic estimate of the
effective temperature, \teff $= 5900 \pm 100$~K, provides an intrinsic
bolometric luminosity of $1.9 < L_\star/L_\odot < 7.6$.

In Table~\ref{tbl3}, we show distance determinations for the different
filters, using the bolometric corrections for models without
overshooting (Bessell, Castelli, \& Plez 1998), 
for our preferred value of the color
excess, $E(B-V)=0.76$ mag, and other different sets of the relevant
parameters. We compute the magnitude corrected for extinction in each
filter as $m_{V,0}=m_V-A_V$, where $A_V = 3.12\, E(B-V)$ is the
extinction in the Johnson $V$ filter.  We adopt the following values
for other filters (Schaifers et al. 1982): $A_R/A_V=0.84$,
$A_J/A_V=0.32$, $A_H/A_V=0.27$, and $A_K/A_V=0.21$.

Table~\ref{tbl3} clearly shows the uncertainty in the distance
due to the uncertainty in the stellar parameters of Tycho G. 
Note that the relatively small formal error in the distance estimate
for each filter is calculated by
assuming that the magnitudes equal $m_i+\Delta m_i$.

In Table~\ref{tbl3}, we also show for each set of parameters the
average distance determinations weighted by the uncertainty of each
individual distance determination. These values cover a distance range
2.48--4.95 kpc with errors of 0.21--0.52 kpc. We have also considered
the possibility that Tycho G lost part of its envelope, $\Delta
M=-0.2$ M$_\odot$, due to the impact of the supernova shock wave (Marietta, 
Burrows, \& Fryxell 2000; Pakmor et al. 2008). However, a 
lower mass does not produce a
significant change in the derived average distance. In all cases, we
have assumed that the companion star is able to almost completely 
recover thermal equilibrium between $\sim102$ and $\sim103$ yr 
after the white dwarf explosion (Podsiadlowski 2003). The 
distance to the SN 1572 remnant inferred from the expansion of the
radio shell and by other methods is $2.83 \pm 0.79$ kpc (Ruiz-Lapuente
 2004). Such a distance is in good agreement with the derived
distances of Tycho G.

\section{Projected Rotational Velocity of Tycho G}

Tycho G is not a fast rotator, unlike the companion stars of 
Type II SNe. The projected rotational velocity is constrained by
the instrumental resolution to $v_{\rm rot}\, {\rm sin} \, i \leq 6.6$ km
s$^{-1}$.  

However, the companion stars of SNe~II and other core-collapse SNe
have different rotational histories from those proposed for SN~Ia
progenitors; they would be massive stars, and maybe fast rotators
(Meynet \& Maeder 2005, and references therein). Moreover, 
the orbital elements just before the explosion are also different; in
particular, the separations are larger. Given the orbital history up
to the explosion, it is reasonable to find fast rotators around black
holes and neutron stars (Gonz\'alez Hern\'andez et al. 2004, 2005,
2008b).
The companion stars of these compact objects
do not transfer mass to the star that becomes a core-collapse
supernova. If the explosion does not unbind the system, the companion
can be seen as a star orbiting a black hole or neutron star and
showing significant rotational velocity. But, for SNe~Ia, we should
not expect such high velocities since the companion of the white dwarf
would have transferred mass and angular momentum to the white
dwarf. And, after the impact, the loss of the envelope would have
further helped decrease the angular momentum and to brake rotation.

Let us consider, for instance, a WD mass $M_{WD} = 1.4$
M$_{\odot}$ and a companion of mass $M = 1$ M$_{\odot}$ (a system
similar to U Scorpii; see, e.g., Thoroughgood et al. 2001). In such a
system, an orbital velocity of 90 \kms (this value could roughly
correspond to the peculiar velocity of Tycho G estimated by
Ruiz-Lapuente et al. 2004, still consistent with the values from the
present work) 
corresponds to an orbital separation $a = 19.28$ R$_{\odot}$ and an
orbital period $P_{\rm orb} = 6.32\ {\rm d}$. The effective Roche
lobe radius of the companion can be calculated from the Eggleton (1983)
approximation (accurate up to a few percent for all mass ratios):

\begin{equation}
R_{L} = a\left[{0.49\over 0.6 + q^{-2/3}\ {\rm ln}(1 + q^{1/3})}\right].
\end{equation}
It would then be $R_{L} = 6.75$ R$_{\odot}$. 

Assuming that tidal interaction had been effective enough to keep the
rotational period of the companion locked with the orbital period in
spite of the angular momentum loss due to mass transfer, and that no
further changes have intervened, the rotation period of the companion
should now be $P_{\rm rot} = P_{\rm orb} = 6.32\ {\rm d}$.  Therefore,
the rotational velocity is $v_{\rm rot} = 2\pi R/P_{\rm rot}$.
However, the present radius of the Tycho G is within the range $R
\approx 1$--2 R$_{\odot}$, as can be seen from its position in the
Hertzsprung-Russell diagram. This then gives $v_{\rm rot} \approx
8$--16 \kmso. Here we have assumed that the impact of the SN ejecta
has removed the outer, more distended and less gravitationally bound
layers (Marietta et al. 2000; Pakmor et al. 2008), and that the star
is able to completely or partially recover thermal equilibrium
$\sim500$ yr after the explosion (Podsiadlowski 2003). In this case,
when the energy deposited on the star due to the impact of the shock
wave of the supernova is $\lesssim 5\times 10^{46}$ erg, the
luminosity of the star in the time interval $\sim102$--$103$ yr after
the explosion could be roughly $\sim 2$--5 \Lsuno. Hence, the radius of
the star would be in the range $\sim1.5$--2 \Rsuno.
 
Thus, an observed value of $v_{\rm rot}\, {\rm sin}\ i \leq 6.6$
\kms is not an unexpected result in this case (for a typical angle $i
\approx 45^{o}$ or lower).  In fact, the preceding analysis does not
even take into account angular momentum losses due to mass stripping,
nor the slowing down of rotation (with conservation of angular
momentum) that would result from puffing up of the fraction of the
envelope remaining bound.

Any suggestion that the rotational velocity of the former companion of
a SN~Ia should now be very high is based on the possibly incorrect 
assumption that its current radius should remain, after the SN~Ia 
explosion, at the large value it had at the time of Roche-lobe filling 
(equal to 6 R$_{\odot}$ in our example). Thus, expected rotational 
velocities of companions of supernovae, as well as potential pollution 
by the supernova ejecta, are different in SNe~Ia and SNe~II/Ib/Ic.

\section{Space Velocity and Position of Tycho G}

In our previous study (Ruiz-Lapuente et al. 2004), we measured the
proper motion of Tycho G to be $\mu_{b} = -6.11 \pm 1.34$ mas
yr$^{-1}$ and $\mu_{l} = -2.6 \pm 1.34$ mas yr$^{-1}$. The proper
motion in Galactic latitude implies that the star is an outlier in
proper motion, with a derived tangential velocity of $94 \pm 27$ \kmso.
A new, more accurate and precise proper motion is currently being
derived from {\it HST} images obtained in Cycle 16.

The spectral data presented by Ruiz-Lapuente et al. (2004) allowed an
estimate of the heliocentric radial velocity of Tycho G with an
uncertainty of 10--18 \kmso. The new Keck HIRES data provide a much
smaller uncertainty. For this new measurement we selected six strong,
unblended lines in the spectral range 6000--8000~\AA\ and measured the
radial velocity of each individual line. The average radial velocities
from those lines are found to be $-87.36 \pm 0.60$ \kms and $-87.53
\pm 0.48$ \kms during the first and second nights, respectively,
where the errors indicate the dispersion of the measurements. The
average of the two epochs is $v_r = -87.4 \pm 0.5$ \kmso.  This value
implies that $v_r \approx -80$ \kms in the local standard of rest
(LSR). In our previous measurement, we had quoted $v_r = -99 \pm 6$ \kms
~(statistical error). The measurement was made using H${\alpha}$ and
H${\beta}$ lines, which are worse indicators of the velocity of the
star.

\begin{figure}[!ht]
%\epsscale{0.70}
%\plotone{f1bw.eps}
\centering
\includegraphics[width=8cm,angle=0]{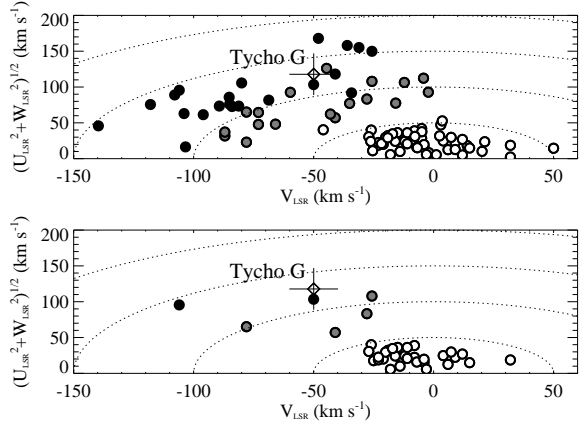}
\caption{Upper panel: Toomre diagram of the thick-disk stars (black 
filled circles), thin-disk stars (open circles), and transition stars 
(grey filled circles) from Bensby et al. (2003, 2005). The star Tycho G is
displayed as a rhombus. Lower panel: the same as upper panel, but for 
stars with metallicity greater than or equal to the metallicity of Tycho G.}   
\label{toomrefig}
\end{figure}

Stars in the direction of the Tycho SN remnant at distances of 2--4
kpc move at an average radial velocity $v_r \approx -20$ to $-40$~\kms
(in the LSR), in agreement with Galactic rotation models (Brand \&
Blitz 1993; Dehnen \& Binney 1998). Tycho G is well above the average
of stars in its vicinity. Both the radial velocity and the proper
motion show a comparably high velocity. One could argue that Tycho G
might be at a larger distance (although this seems unlikely from
its photometric magnitudes; see \S~\ref{dist}), but the observed
proper motion would make the tangential velocity unreasonably high if
the star were $>6$~kpc away ($v_t \approx 187$~\kms at $d=6$~kpc). 
On the other hand,
a distance of 4--5~kpc (although $v_t \approx 140$~\kms at $d=4.5$~kpc)
might not be ruled out by
these arguments, albeit somewhat inconsistent with the estimated
distance of the Tycho SN 1572 remnant ($2.83 \pm 0.79$~kpc;
Ruiz-Lapuente 2004). 

However, Fuhrmann (2005) has argued, based on the Galactic space 
velocity components of Tycho G relative
to the LSR ($U_{\rm LSR},V_{\rm LSR},W_{\rm LSR}$),
that this star could be just a thick-disk star passing close to 
the SN remnant. Assuming that the Sun
moves with components $(U_{\odot,{\rm LSR}},V_{\odot,{\rm LSR}},W_{\odot,{\rm
LSR}})=(10.,5.2,7.2)$ \kms (Dehnen \& Binney 1998)
relative to the LSR, for distance of 3\,kpc, we estimate 
the space Galactic velocity components of Tycho G to be 
$(U_{\rm LSR},V_{\rm LSR},W_{\rm LSR})= (85\pm17,-50\pm10,-82\pm24)$ \kmso.
In Figure~\ref{toomrefig} we display Galactic velocity of Tycho G in 
comparison with the Galactic velocities of thick-disk and thin-disk 
stars. Although the Galactic velocity of Tycho G is comparable to 
that of thick-disk stars, its metallicity is more common in thin-disk 
stars. In the lower panel of Figure~\ref{toomrefig}, we display only those 
stars with metallicity equal to or greater than the metallicity of Tycho G.
The number of thick-disk stars is reduced considerably, but there are still 
a few thick-disk stars with the same metallicity and Galactic velocity as
Tycho G. On the other hand, the high Ni abundance of Tycho G would be 
very unusual for a thick-disk star. We therefore conclude that while a
thick-disk origin may be consistent with the data, this possibility
is not favored because it provides no explanation for the anomalous 
Ni abundance. 

Although the position of Tycho G in the sky does not coincide with the 
proposed geometrical center of the X-ray and radio remnants of SN 1572, 
the offset of the position is within the uncertainties due to 
the asymmetry resulting from interaction of the expanding remnant 
with the circumstellar medium. Indeed, there is $0.56'$ displacement 
along the E--W axis between the radio emission and the high-energy 
continuum observed by XMM-Newton, in the position of the western 
rim (Decourchelle et al. 2001). Such asymmetry amounts to a 14\% 
offset along the E--W axis. It appears to be due to the ejecta 
having encountered a dense H cloud at the eastern edge, giving rise to
brighter emission and lower expansion velocity of the ejecta there,
while interacting with a lower-density medium in the western rim. It
was precisely owing to those considerations that we studied all of the
23 stars with $m_{V} \le 22$ mag
within a circle of radius $0.65'$ around the geometrical center
of the X-ray emission as given by the Chandra observations (Hughes
2000). All of the stars within that circle were considered on an equal
footing as possible companions until all were discarded with the
exception of Tycho G, due to the unique characteristics of this star.
In the next two sections, we discuss its chemical peculiarities.

\begin{deluxetable}{lrrrrrrrrr}
\tabletypesize{\scriptsize}
\tablecaption{Supernova Ia models in \mbox{Tycho G}\label{tbl4}}
\tablewidth{0pt}
\tablehead{ & & & \multicolumn{7}{c}{${\rm [X/H]\:EXPECTED}\tablenotemark{c}$} \\
\tableline
\colhead{ELEMENT} & \colhead{${\rm [X/H]\:\rm OBSERVED}\tablenotemark{a}$} & 
\colhead{${\rm [X/H]}_{0}\tablenotemark{b}$} & 
\colhead{W7} & \colhead{W70} & \colhead{WDD1} & \colhead{WDD2} & 
\colhead{WDD3} & \colhead{CDD1} & \colhead{CDD2}}            
\startdata
      O  &   $-0.02 \pm  0.11$ & $-$0.15 & $-$0.14 & $-$0.14 & $-$0.14 & $-$0.15 & $-$0.15 & $-$0.14 & $-$0.15 \\ 
     Na  &   $ 0.04 \pm  0.12$ & $-$0.26 & $-$0.26 & $-$0.26 & $-$0.26 & $-$0.26 & $-$0.26 & $-$0.26 & $-$0.26 \\ 
     Mg  &   $ 0.08 \pm  0.08$ & $-$0.10 &  0.04 &  0.13 &  0.03 & $-$0.02 & $-$0.05 &  0.03 & $-$0.02 \\ 
     Al  &   $ 0.16 \pm  0.07$ & $-$0.06 & $-$0.05 & $-$0.06 & $-$0.06 & $-$0.06 & $-$0.06 & $-$0.06 & $-$0.06 \\ 
     Si  &   $-0.03 \pm  0.05$ & $-$0.22 & $-$0.07 & $-$0.08 &  0.02 & $-$0.03 & $-$0.06 &  0.02 & $-$0.03 \\ 
      S  &   $-0.02 \pm  0.13$ & $-$0.26 & $-$0.10 & $-$0.10 & $-$0.00 & $-$0.05 & $-$0.09 &  0.00 & $-$0.06 \\ 
     Ca  &   $-0.04 \pm  0.11$ & $-$0.31 & $-$0.15 & $-$0.08 &  0.03 & $-$0.03 & $-$0.08 &  0.03 & $-$0.03 \\ 
     Ti  &   $-0.03 \pm  0.12$ & $-$0.06 &  0.00 &  0.02 &  0.12 &  0.10 &  0.09 &  0.08 &  0.07 \\ 
     Cr  &   $-0.04 \pm  0.09$ & $-$0.27 &  0.03 &  0.05 &  0.27 &  0.21 &  0.16 &  0.24 &  0.20 \\ 
     Mn  &   $ 0.02 \pm  0.12$ & $-$0.27 &  0.12 &  0.05 &  0.11 &  0.06 &  0.03 &  0.09 &  0.05 \\ 
     Fe  &   $-0.05 \pm  0.09$ & $-$0.21 &  0.11 &  0.12 &  0.09 &  0.13 &  0.15 &  0.08 &  0.14 \\ 
     Co  &   $ 0.13 \pm  0.12$ & $-$0.14 &  0.03 &  0.02 & $-$0.07 & $-$0.03 & $-$0.01 & $-$0.09 & $-$0.03 \\ 
     Ni  &   $ 0.11 \pm  0.10$ & $-$0.27 &  0.40 &  0.32 &  0.06 &  0.16 &  0.23 &  0.04 &  0.18 \\ 
\enddata
\tablenotetext{a}{Observed abundances of the secondary star in \mbox{Tycho G}.}

\tablenotetext{b}{Initial abundances assumed for
the secondary star in \mbox{Tycho G}; see text.}

\tablenotetext{c}{Expected abundances in Tycho G after being
contaminated with nucleosynthetic products of SN~Ia explosion
models from Iwamoto et al. (1999). Some of them are presented
in Fig.~\ref{snfig}.}

\end{deluxetable}

\section{Discussion}

\subsection{Has Tycho G been Polluted by the SN Ia?\label{polluted}}

The measured ratio [Ni/Fe] $= 0.16 \pm 0.04$ in Tycho G is 4--5$\sigma$
above the average value of this ratio in the Galaxy ([Ni/Fe] $\approx
-0.05$). It indicates an overabundance of Ni with
respect to the Galactic trend. Ni may have been trapped by the star as
slowly moving material from deep layers of the SN ejecta 
orbited around it. 
While the rapidly moving ejecta consisting mainly of intermediate-mass
elements would have escaped away from the site of the explosion, the
very slowly moving tail made of heavy elements like Ni could have been 
captured by the companion.

To be consistent with the evolutionary path suggested above (and also
by Ruiz-Lapuente et al. 2004), we use $a = 19.28$ R$_{\odot}$ for the
separation between the white dwarf and its companion, and 
$R_{*}= R_{L} = 6.75$ R$_{\odot}$ for the radius of the companion 
(just before the SN explosion).

The mass that could be trapped by the companion is given by the
subtended angle,

\begin{equation}
m_{\rm t} = \Delta M  \ (\pi R_{2}^{2}/4\pi \ a^{2})  \ f, 
\end{equation}
 
\noindent
where $\Delta M$ is  the ejected mass and $f$ is the fraction
of retained material.  
 
\noindent

We try to model the contamination of the companion star by the nucleosynthetic
products of different SN~Ia models from Iwamoto et al. (1999). In
Table~\ref{tbl4} we show the expected abundances of Tycho G after having
captured a significant amount of the SN ejecta. In these model computations, 
we adopt the yields from SN~Ia models with different kinetic explosion 
energies in the range (1.30--$1.44)\,\times\,10^{51}$ ergs, different 
central densities (in units of $109$ g~cm$^{-3}$) of $\rho_9=1.37$
(C) and 2.12 (W), and different deflagration velocities with
fast deflagration in the models W7 and W70 (Nomoto, Thielemann, \&
Yokoi 1984) and slow deflagration in the
others. We assumed as initial abundances the average abundances of disk stars
with a metallicity of ${\rm [Fe/H]} = -0.20\pm 0.09$, and a capture
efficiency factor $f=0.03$. The matter that is captured by the companion has a
much larger mean molecular weight than the composition of its atmosphere and
then is completely mixed with the whole star due to thermohaline mixing, in a
short timescale (Podsiadlowski et al. 2002). 

Badenes et al. (2006) found, by comparing a grid of X-ray
synthetic spectra based on hydrodynamical models, that the fundamental
properties of X-ray emission in Tycho SN 1572 are well reproduced
with one-dimensional delayed detonation models having a kinetic energy
of $\sim1.2\,\times\,10^{51}$ ergs. In Table~\ref{tbl4}, the models
called WDD1,2,3 and CDD1,2 include transitions
from slow deflagration to detonation, which are equivalent to
delayed detonation models.

\begin{figure}[!ht]
%\epsscale{0.70}
%\plotone{f1bw.eps}
\centering
\includegraphics[width=8cm,angle=0]{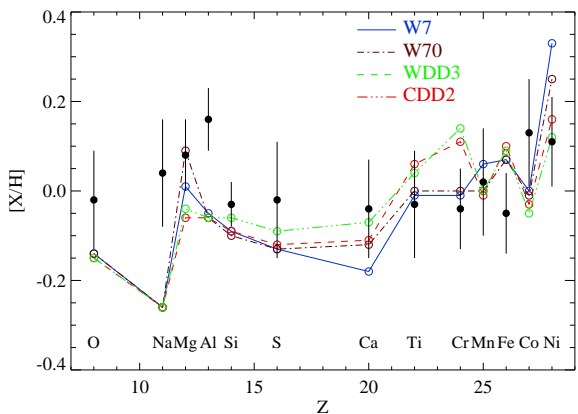}
\caption{Expected abundances in Tycho G after contamination from the
nucleosynthetic products in SN~Ia models (Iwamoto et al. 1999), in
comparison with the observed abundances.}   
\label{snfig}
\end{figure}

In Figure~\ref{snfig}, we display the observed abundances of Tycho G in 
comparison with the expected abundances from several SN~Ia models. 
Except for Na and Al, we find reasonable
agreement between the observed abundances of Tycho G and the expected
abundances after contamination from the nucleosynthetic products of
the SN~Ia. Surprinsingly, Gonz\'alez Hern\'andez et al. (2008a)
measured the chemical abundances of the secondary star of
the black hole binary Nova Scorpii 1994 and found several
$\alpha$ elements significantly enhanced in the companion star while Al
and Na were not enhanced. 
The enhanced $\alpha$ elements in that star were explained by
comparing the observed abundances with supernova/hypernova explosion
models. However, these models also predict that Al and Na should be
enhanced at a comparable level as oxygen, and they did not find an
explanation for this discrepancy. 
Therefore, the model of contamination of Tycho G by using SN~Ia models
could provide an explanation for the relatively high observed [Ni/Fe]
with respect to the Galactic trend of this element. 

One could find stable isotopes of Co as well among the slowly moving
material in a SN~Ia. The ratio [Co/Fe] $ = 0.18 \pm 0.08$ in Tycho G is
consistent with contamination from the SN ejecta. The study by Reddy
et al. (2006) finds that [Co/Fe] shows flat behavior with
[Fe/H] for thin-disk stars ([Co/Fe] $ = -0.05 \pm 0.02$). For
thick-disk stars, a weak trend of decreasing [Co/Fe] with increasing
[Fe/H] is apparent, falling to [Co/Fe] $ = 0.00 \pm 0.04$ at the
metallicity of Tycho G. The effect of a larger [Co/Fe] is $>1\sigma$, 
a lower significance level compared to the ratio [Ni/Fe] (see also 
Fig. 6).  The ratios of the $\alpha$ elements (O, Si, Ca, and Ti) to 
iron do not show enhancements over Galactic values.

\begin{figure*}[!ht]
%\epsscale{0.70}
%\plotone{f1bw.eps}
\centering
\includegraphics[width=15cm,angle=0]{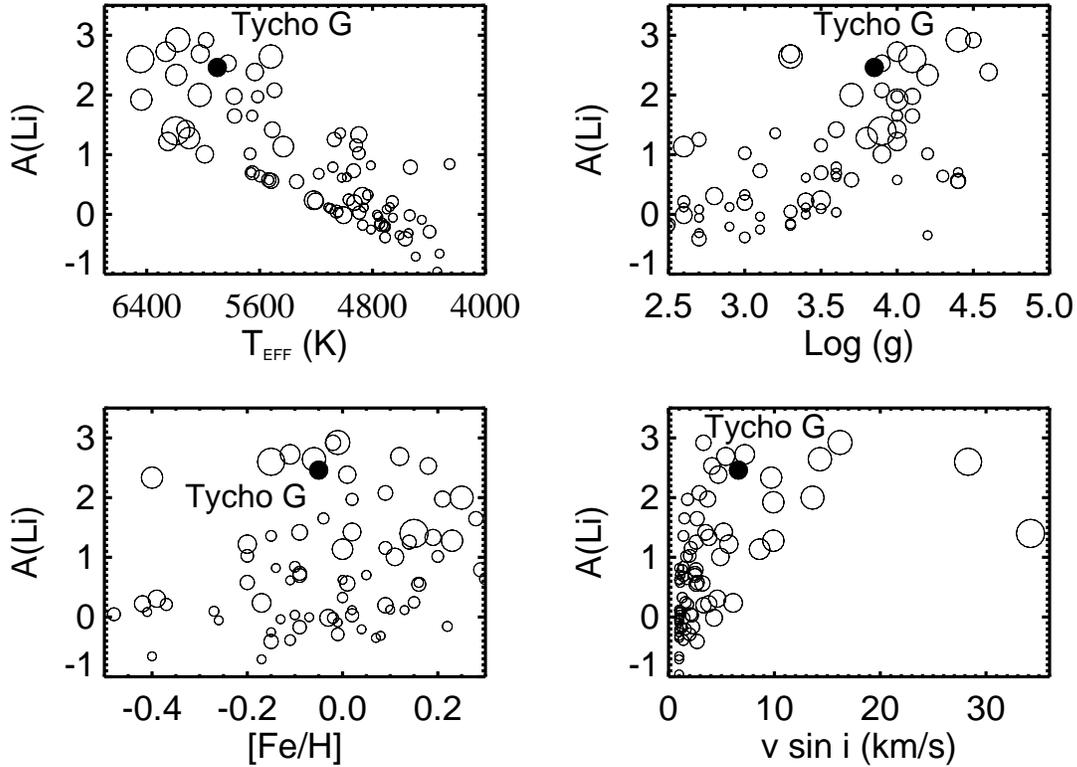}
\caption{Li abundance of Tycho G in comparison with that of subgiant
stars from Randich et al. (1999) having similar stellar parameters,
metallicity, and rotational velocity.}   
\label{lifig}
\end{figure*}

A study of the SN 1572 remnant in X-rays (Hamilton et al. 1985)
suggests that it contains $\sim$0.66 \Msuno, including 0.20 \Msun
of Fe and 0.02 \Msun of Ni in an inner layer as well as $\sim$0.44 \Msun 
of Fe plus Ni unshocked by the reverse shock and consequently moving in
free expansion. The energy of this supernova is estimated from the
ejecta to be $E \approx$ (4--5) $\times 10^{50}$ ergs (Hughes 2000).

\subsection{Li Abundance\label{licap}}

As shown in Figure~\ref{fig3}, the Li line at 6708~\AA\ is pronounced in
Tycho G.  The abundance of this element, $A$(Li) = $2.50 \pm 0.09$,
raises the question of how much Li could have survived in this star
that has already evolved off the main sequence. Similar abundances
have been found in some turn-off stars (with $6300 < T_{\rm eff} <
6800$~K) in the sample of do Nascimento et al. (2000, 2003), and in a
few subgiant stars with stellar parameters similar to those of Tycho
G. In the sample of Randich et al. (1999), there are also a few
subgiants with similar stellar parameters and Li abundance as Tycho G;
see Figure~\ref{lifig}. 
Although the Li abundance of Tycho G is somewhat high for its relatively 
low effective temperature and surface gravity, the upper-right panel 
of Figure~\ref{lifig} shows two stars (with \teff $\approx 5500$ K and
6000~K, and \loggl\ $\approx 3.3$ dex) that are even more anomalous than
Tycho G.

Canto Martins et al. (2006) found a Li-rich subgiant star (S1242) in
the open cluster M67, with \teff $=5800$~K, \loggl\ $=3.9$ dex, [Fe/H]
$= -0.05$, and $A$(Li) = 2.7, all similar to the properties of Tycho
G. This star is in a wide eccentric binary system, and these authors
proposed that the Li has been preserved due to tidal effects. There
exist other subgiant stars of similar type showing high Li abundances
and belonging to binary systems. Again, this could be linked to being
in a close binary system where tidal interaction might inhibit Li
destruction.

An alternative is that Li was created as a result of energetic
processes. Mart{\'\i}n et al. (1992, 1994a,b) suggest that in soft
X-ray transients (SXTs) one is observing freshly synthesized Li, and
that the Li could come from $\alpha \alpha$ and/or other spallation
reactions during the outbursts of these binary systems that contain
black holes or neutron stars.  Spallation reactions are produced by
protons and $\alpha$ particles with energies above a few MeV when
hitting C, N, and O nuclei, and also by $\alpha + \alpha$ collisions
at similar energies.

In addition, in the case where the compact object is a white dwarf,
Mart{\'\i}n et al. (1995) have shown that their companion stars do not
show high Li abundances. They claimed that this is consistent with the
Li production scenario in SXTs, because the weaker gravitational
potential well of the white dwarfs does not allow particles to be
accelerated to the required high energies (Mart{\'\i}n, Spruit, \&
van Paradijs 1994b).
However, other authors have shown that Li production is possible
during nova explosions (Starrfield et al. 1978; Boffin, Paulus, \&
Arnould 1993).

The energy arguments could also work for the enhancement of Li in the
companion of a carbon-oxygen white dwarf. The accretion lasts for a
sufficiently long time to build enough Li nuclei in the convective
envelope of the secondary star. At the onset of the explosion, the
supernova is able to unbind part of the material from the convective
envelope. If part of the convective envelope were to survive, it would
display a high Li abundance.  However, this scenario of Li production
during outbursts in SXTs also predicts Li isotopic ratios in the range
$0.1 < N(^6$Li)/$N(^7$Li$) < 10$; see Casares et al. (2007), and
references therein. These authors conclude that the preservation
scenario by tidal effects is the most likely mechanism to explain the
high Li abundances of SXTs.  

There are two other alternatives for enrichment of Li in the
companions of SNe~Ia. The first is that the high-energy particles
accelerated by the shock wave in the outermost layers of the exploding
white dwarf induce spallation reactions in the surface of the
companion. A fraction of the irradiated material could mix with the
layers now making the surface of the companion, and Li enrichment
would show.  The second alternative is that the surface of the
companion, after the explosion, can be bombarded by high-energy
particles accelerated in the SN remnant, producing spallation
reactions. Steady bombardment by those high-energy particles has the
restriction that while it enriches Li, the energy deposited would
increase the star's luminosity.  However, if Li enrichment is limited
to a thin layer extending not much below the photosphere, the increase
in luminosity would be negligible.

\section{Overall Discussion of Tycho G}

Ruiz-Lapuente et al. (2004) suggested Tycho G as the mass-donating
companion of SN 1572 and gave stellar parameters that classified it as
a G0--G2~IV star. In the present study, higher-resolution spectra have
enabled us to pin down the stellar parameters: $T_{\rm eff} = 5900 \pm
100$ K, \loggl\ $= 3.85 \pm 0.30$ dex, and [Fe/H] $= -0.05 \pm
0.09$. Thus, we have been able to confirm that Tycho G has a
metallicity close to solar and its luminosity class is that of a
subgiant near the main sequence. Recently, Badenes et al. (2008) 
have estimated the global metallicity of the Tycho SN~Ia remnant at
$\log(Z_{\rm SN}/Z_\odot) = 0.45^{+0.31}_{-0.60}$ when adopting the
solar abundances of Grevesse \& Sauval (1998). We can estimate the
global metallicity of the star by adding all the values of
[X/H] provided in Table~\ref{tbl2}. The metallicity of Tycho G would be 
$\log(Z_{\rm Tycho G}/Z_\odot) \approx 0.23$, which is consistent with the
value given by Badenes et al. (2008) within their large error bars.

Moreover, in support of the hypothesis that Tycho G is physically
within the remnant of SN 1572 and is possibly the surviving companion of 
the supernova, we have found a high value of [Ni/Fe], $0.16 \pm 0.04$, 
about 3$\sigma$ above the average value in Galactic disk
stars ([Ni/Fe] $\approx -0.05\pm0.03$). In addition, we report an
anomalously high Li abundance for a subgiant, $A$(Li) $= 2.50 \pm
0.09$, although similar to a few other known subgiant stars having
similar stellar parameters and metallicity. 

The star is moving at high speed both in radial velocity and proper
motion. If we consider the possibility that Tycho G is actually beyond
the supernova remnant, at 6 kpc or more, where $v_{r,\rm LSR} = -80$
\kms would follow the low-latitude rotation pattern of the Galaxy, we
run into two other physical incompatibilities: (a) at 6 kpc in the
direction of SN 1572, the region has metallicities [Fe/H] $< -0.3$,
much lower than observed for Tycho G; and (b) the tangential velocity
implied by the observed proper motion and distance would be
unreasonably high ($v_t \approx 187$ \kms at $d=6$ kpc). On the other
hand, according to its space Galactic velocity components, Tycho G
could be just a thick-disk star passing near the SN~Ia remnant, but 
its high Ni abundance argues against this possibility.

We find that Tycho G has features in common with the companions of
black holes and/or neutron stars that presumably originated in Type II
SNe. These companions, which remain bound to the black hole or neutron
star created by the explosion, are main-sequence stars and subgiants
with a high Li abundance. The companions of core-collapse SNe are
polluted with $\alpha$ elements, but Tycho G has an overabundance of
Ni (and Co at lower significance level) with respect to Fe. The
overabundance is well above the scatter for this ratio in Galactic
disk stars, suggesting that Tycho G could have captured the
low-velocity tail of the SN 1572 ejecta. We use SN~Ia yields to model
the possible contamination of the star from the SN ejecta and we find
reasonable agreement with the observed abundances of Tycho G.

Thus, although the current evidence is not fully compelling,
it remains consistent with Tycho G being the surviving companion 
star of the white dwarf that exploded to form SN 1572, as
originally proposed by Ruiz-Lapuente et al. (2004).

\acknowledgments

J.I.G.H. is supported by EU contract MEXT-CT-2004-014265
(CIFIST). P.R.-L. acknowledges support from AYA2006-05639.  A.V.F. is
funded by NSF grant AST--0607485, as well as by NASA/{\it HST}
grant GO--11114 from the Space Telescope Science Institute,
which is operated by the Association of
Universities for Research in Astronomy, Inc., under NASA contract
NAS5--26555. We are grateful to Tom Marsh for the
use of the MOLLY analysis package, and to Ryan Chornock for very
helpful discussions on a number of issues. P.R.-L. is grateful to 
Eduardo Mart{\'\i}n and Rafa Rebolo for useful exchanges on the Li topic. 
A.G. acknowledges support by the Benoziyo Center for Astrophysics 
and the William Z. and Eda Bess Novick New Scientists Fund at the
Weizmann Institute. This study is based primarily on data from
the W. M. Keck Observatory, which is operated as a scientific 
partnership among the California Institute of
Technology, the University of California, and NASA; it was made
possible by the generous financial support of the W. M. Keck
Foundation. We thank the Keck staff for their assistance with the
observations. This work has utilized IRAF facilities and the 2MASS
catalog.

%\vfill\eject

\appendix

{\bf Appendices}

\section{Low-Resolution Spectra\label{ap1}}

\begin{figure*}[!ht]
%\epsscale{0.70}
%\plotone{f1bw.eps}
\centering
\includegraphics[width=12.cm,angle=0]{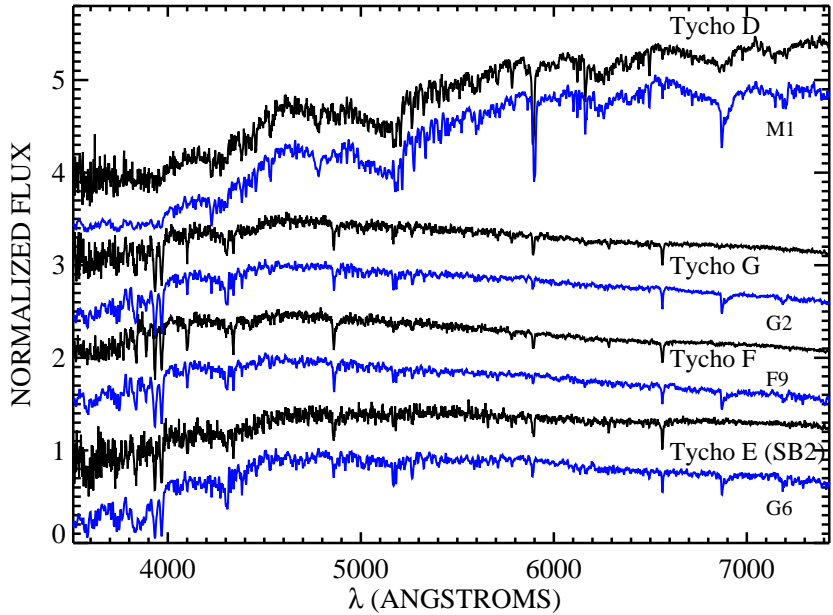}
\caption{LRIS spectra of four targets of the Tycho SN 1572 field
compared with template low-resolution spectra from the Jacoby et
al. (1984) library. Note that telluric lines (B band near 6900~\AA;
weak lines around 7200~\AA) were not removed from the templates. The
spectra of Tycho G and F were dereddened with $E(B-V)=0.76$ mag,
whereas $E(B-V)=0.66$ mag was used for the spectra of Tycho D and
E. The best fit to each spectrum is also shown. Note that when Tycho E
is observed at high resolution it seems to be a double-lined
spectroscopic binary, and therefore the fit shown is not reliable.}
\label{lrall}
\end{figure*}

We attempted to determine the spectral type of several stars in the
Tycho SN 1572 field. In Figure~\ref{lrall} we display the best fits to
the LRIS spectra of Tycho D, E, F, and G (see Ruiz-Lapuente et
al. 2004 for identifications). The spectra of Tycho G and F were
dereddened with $E(B-V)=0.76$ mag, whereas the spectrum of Tycho D
was dereddened with $E(B-V)=0.66$ mag since this color excess
provides a better fit to the LRIS spectrum. Based on the closer
position of Tycho D and E to the center of the Tycho SN 1572 field, we
also applied a color excess of $E(B-V)=0.66$ mag to Tycho E.  However,
we know from high-resolution spectra that Tycho E is in fact a
double-lined spectroscopic binary, so this comparison using a
low-resolution composite spectrum is meaningless. The spectral
classifications of most stars differ from those published by Ihara et
al. (2007), but are consistent with the results of Ruiz-Lapuente
et al. (2004).

\section{Line List with Equivalent Widths for Tycho G}

Here we provide the complete list of line EWs measured 
from the high-resolution Keck spectra of Tycho G. 

\begin{deluxetable}{lrrrrrrrrr}
\tabletypesize{\scriptsize}
\tablewidth{0pt}
\tablecaption{Equivalent Widths of Lines in Tycho G\label{tblB}}
\tablehead{\colhead{Species} & \colhead{$\lambda$} & \colhead{$\chi$}  &
\colhead{$\log (gf)$} & \colhead{EW$_\lambda$} & \colhead{Species} &
\colhead{$\lambda$} & \colhead{$\chi$}  & \colhead{$\log (gf)$} &
\colhead{EW$_\lambda$} \\ 
 & \colhead{(\AA)} & \colhead{(eV)} &  &  \colhead{(m\AA)} & &
 \colhead{(\AA)} & \colhead{(eV)} &  &  \colhead{(m\AA)} } 
\startdata
\ion{Li}{1}  & 6707.81 &  0.00 &  0.175 & $ 38.7 \pm  3.9$ &      \ion{Mn}{1} & 4502.220 & 2.920 & -0.490 & $ 60.3 \pm 12.3$ \\ 
  &  &  &  &  & 						  \ion{Mn}{1} & 5394.670 & 0.000 & -3.500 & $ 51.9 \pm  4.7$ \\ 
\ion{C}{1}  & 5052.160 & 7.680 & -1.420 & $ 38.7 \pm  7.0$ &  	  \ion{Mn}{1} & 5399.470 & 3.850 & -0.097 & $ 32.0 \pm  4.7$ \\ 
\ion{C}{1}  & 6587.620 & 8.540 & -1.000 & $ 16.5 \pm  3.8$ & 	  \ion{Mn}{1} & 5420.360 & 2.140 & -1.460 & $ 38.5 \pm  4.6$ \\ 
\ion{C}{1}  & 7113.170 & 8.650 & -0.770 & $ 23.6 \pm  3.3$ & 	  \ion{Mn}{1} & 5432.540 & 0.000 & -3.620 & $ 36.0 \pm  4.5$ \\ 
\ion{C}{1}  & 7115.170 & 8.640 & -0.930 & $ 32.2 \pm  3.3$ & 	  \ion{Mn}{1} & 6021.800 & 3.070 &  0.030 & $ 91.0 \pm  4.4$ \\ 
\ion{C}{1}  & 7116.960 & 8.650 & -0.910 & $ 25.3 \pm  3.3$ & 	    &  &  &  &   \\
  &  &  &  &  & 						  \ion{Fe}{1}  & 5044.220 & 2.850 & -2.040 & $ 75.9 \pm  7.0$ \\ 
\ion{O}{1}  & 7771.960 & 9.110 &  2.831 & $ 94.7 \pm  2.8$ & 	  \ion{Fe}{1}  & 5247.060 & 0.090 & -4.930 & $ 58.0 \pm  5.5$ \\ 
\ion{O}{1}  & 7774.180 & 9.110 &  2.061 & $ 82.3 \pm  2.8$ & 	  \ion{Fe}{1}  & 5322.050 & 2.280 & -2.900 & $ 47.6 \pm  5.1$ \\ 
\ion{O}{1}  & 7775.400 & 9.110 &  1.256 & $ 68.0 \pm  2.8$ & 	  \ion{Fe}{1}  & 5806.730 & 4.610 & -0.890 & $ 50.3 \pm  4.4$ \\ 
  &  &  &  &  & 						  \ion{Fe}{1}  & 5852.220 & 4.550 & -1.190 & $ 42.9 \pm  4.4$ \\ 
\ion{Na}{1}  & 5688.220 & 2.104 & -0.625 & $117.4 \pm  4.3$ & 	  \ion{Fe}{1}  & 5855.080 & 4.610 & -1.530 & $ 12.4 \pm  4.4$ \\ 
\ion{Na}{1}  & 6154.230 & 2.102 & -1.607 & $ 34.7 \pm  3.8$ & 	  \ion{Fe}{1}  & 5856.090 & 4.290 & -1.560 & $ 23.5 \pm  4.4$ \\ 
\ion{Na}{1}  & 6160.750 & 2.104 & -1.316 & $ 40.1 \pm  3.8$ & 	  \ion{Fe}{1}  & 6027.060 & 4.080 & -1.180 & $ 56.4 \pm  4.4$ \\ 
  &  &  &  &  & 						  \ion{Fe}{1}  & 6056.010 & 4.730 & -0.500 & $ 52.7 \pm  4.3$ \\ 
\ion{Mg}{1}  & 4730.040 & 4.346 & -2.390 & $ 52.3 \pm  9.4$ & 	  \ion{Fe}{1}  & 6079.010 & 4.650 & -1.010 & $ 30.6 \pm  4.2$ \\ 
\ion{Mg}{1}  & 5711.090 & 4.346 & -1.706 & $109.1 \pm  4.3$ & 	  \ion{Fe}{1}  & 6151.620 & 2.180 & -3.300 & $ 46.0 \pm  3.8$ \\ 
\ion{Mg}{1}  & 6318.720 & 5.108 & -1.996 & $ 43.4 \pm  3.6$ & 	  \ion{Fe}{1}  & 6157.730 & 4.070 & -1.240 & $ 52.7 \pm  3.8$ \\ 
\ion{Mg}{1}  & 6319.240 & 5.108 & -2.179 & $ 29.3 \pm  3.6$ & 	  \ion{Fe}{1}  & 6165.360 & 4.140 & -1.500 & $ 45.8 \pm  3.8$ \\ 
  &  &  &  &  & 						  \ion{Fe}{1}  & 6180.210 & 2.730 & -2.640 & $ 46.1 \pm  3.7$ \\ 
\ion{Al}{1}  & 6696.030 & 3.143 & -1.570 & $ 37.8 \pm  4.0$ & 	  \ion{Fe}{1}  & 6188.000 & 3.940 & -1.630 & $ 43.5 \pm  3.7$ \\ 
\ion{Al}{1}  & 6698.670 & 3.143 & -1.879 & $ 26.1 \pm  4.0$ & 	  \ion{Fe}{1}  & 6200.320 & 2.610 & -2.400 & $ 73.0 \pm  3.6$ \\ 
  &  &  &  &  & 						  \ion{Fe}{1}  & 6226.740 & 3.880 & -2.070 & $ 33.4 \pm  3.6$ \\ 
\ion{Si}{1}  & 5665.560 & 4.920 & -1.980 & $ 47.0 \pm  4.3$ & 	  \ion{Fe}{1}  & 6229.240 & 2.840 & -2.890 & $ 27.1 \pm  3.6$ \\ 
\ion{Si}{1}  & 5690.430 & 4.930 & -1.790 & $ 52.3 \pm  4.3$ & 	  \ion{Fe}{1}  & 6240.650 & 2.220 & -3.290 & $ 42.3 \pm  3.6$ \\ 
\ion{Si}{1}  & 5701.100 & 4.930 & -2.020 & $ 32.2 \pm  4.3$ & 	  \ion{Fe}{1}  & 6265.140 & 2.180 & -2.560 & $ 71.6 \pm  3.6$ \\ 
\ion{Si}{1}  & 5772.140 & 5.080 & -1.620 & $ 53.0 \pm  4.4$ & 	  \ion{Fe}{1}  & 6270.230 & 2.860 & -2.580 & $ 41.4 \pm  3.6$ \\ 
\ion{Si}{1}  & 5793.090 & 4.930 & -1.910 & $ 44.9 \pm  4.4$ & 	  \ion{Fe}{1}  & 6392.540 & 2.280 & -3.930 & $ 11.8 \pm  3.7$ \\ 
\ion{Si}{1}  & 5948.550 & 5.080 & -1.110 & $ 82.6 \pm  4.5$ & 	  \ion{Fe}{1}  & 6498.940 & 0.960 & -4.630 & $ 37.1 \pm  3.7$ \\ 
\ion{Si}{1}  & 6125.020 & 5.610 & -1.520 & $ 38.9 \pm  3.9$ & 	  \ion{Fe}{1}  & 6608.030 & 2.280 & -3.960 & $ 17.9 \pm  3.9$ \\ 
\ion{Si}{1}  & 6142.490 & 5.620 & -1.480 & $ 36.9 \pm  3.9$ & 	  \ion{Fe}{1}  & 6627.550 & 4.550 & -1.480 & $ 24.7 \pm  3.9$ \\ 
\ion{Si}{1}  & 6145.020 & 5.610 & -1.400 & $ 35.2 \pm  3.9$ & 	  \ion{Fe}{1}  & 6703.570 & 2.760 & -3.020 & $ 29.6 \pm  4.0$ \\ 
\ion{Si}{1}  & 6155.150 & 5.620 & -0.750 & $ 73.2 \pm  3.8$ & 	  \ion{Fe}{1}  & 6710.320 & 1.480 & -4.820 & $ 15.4 \pm  3.9$ \\ 
\ion{Si}{1}  & 6244.480 & 5.610 & -1.360 & $ 33.8 \pm  3.6$ & 	  \ion{Fe}{1}  & 6725.360 & 4.100 & -2.200 & $ 17.7 \pm  3.9$ \\ 
\ion{Si}{1}  & 6721.860 & 5.860 & -1.090 & $ 40.0 \pm  3.9$ & 	  \ion{Fe}{1}  & 6726.670 & 4.610 & -1.050 & $ 33.0 \pm  3.9$ \\ 
  &  &  &  &  & 						  \ion{Fe}{1}  & 6733.160 & 4.640 & -1.430 & $ 25.0 \pm  3.9$ \\ 
\ion{Si}{2}  & 6371.360 & 8.120 & -0.050 & $ 42.7 \pm  3.7$ & 	  \ion{Fe}{1}  & 6750.160 & 2.420 & -2.610 & $ 60.2 \pm  3.9$ \\ 
  &  &  &  &  & 						  \ion{Fe}{1}  & 6752.710 & 4.640 & -1.230 & $ 33.1 \pm  3.9$ \\ 
\ion{S}{1}  & 6046.020 & 7.870 & -0.510 & $ 18.1 \pm  4.3$ & 	    &  &  &  &   \\
\ion{S}{1}  & 6052.670 & 7.870 & -0.630 & $ 12.5 \pm  4.3$ & 	  \ion{Fe}{2} & 5234.630 & 3.220 & -2.230 & $ 98.1 \pm  5.5$ \\ 
\ion{S}{1}  & 6743.440 & 7.870 & -1.270 & $  8.4 \pm  3.9$ & 	  \ion{Fe}{2} & 5425.260 & 3.200 & -3.160 & $ 66.8 \pm  4.6$ \\ 
\ion{S}{1}  & 6757.007 & 7.870 & -0.810 & $ 17.8 \pm  3.9$ & 	  \ion{Fe}{2} & 5991.380 & 3.150 & -3.530 & $ 37.1 \pm  4.5$ \\ 
  &  &  &  &  & 						  \ion{Fe}{2} & 6084.110 & 3.200 & -3.780 & $ 29.5 \pm  4.1$ \\ 
\ion{Ca}{1}  & 5581.970 & 2.520 & -0.650 & $ 93.1 \pm  4.2$ & 	  \ion{Fe}{2} & 6149.250 & 3.890 & -2.720 & $ 49.7 \pm  3.8$ \\ 
\ion{Ca}{1}  & 5590.120 & 2.520 & -0.710 & $ 79.2 \pm  4.3$ & 	  \ion{Fe}{2} & 6247.560 & 3.890 & -2.350 & $ 66.4 \pm  3.6$ \\ 
\ion{Ca}{1}  & 5867.570 & 2.930 & -1.570 & $ 15.5 \pm  4.4$ & 	  \ion{Fe}{2} & 6369.460 & 2.890 & -4.130 & $ 29.9 \pm  3.7$ \\ 
\ion{Ca}{1}  & 6161.290 & 2.520 & -1.220 & $ 58.5 \pm  3.8$ & 	  \ion{Fe}{2} & 6432.690 & 2.890 & -3.560 & $ 61.1 \pm  3.7$ \\ 
\ion{Ca}{1}  & 6166.440 & 2.520 & -1.120 & $ 65.4 \pm  3.8$ & 	  \ion{Fe}{2} & 7479.700 & 3.890 & -3.590 & $ 16.7 \pm  3.0$ \\ 
\ion{Ca}{1}  & 6169.050 & 2.520 & -0.730 & $ 91.5 \pm  3.7$ & 	  \ion{Fe}{2} & 7711.730 & 3.900 & -2.550 & $ 57.9 \pm  2.8$ \\ 
\ion{Ca}{1}  & 6169.560 & 2.520 & -0.440 & $ 98.4 \pm  3.7$ & 	    &  &  &  &   \\
\ion{Ca}{1}  & 6449.820 & 2.520 & -0.630 & $ 99.4 \pm  3.7$ & 	  \ion{Co}{1}  & 4792.860 & 3.250 & -0.150 & $ 37.7 \pm  8.6$ \\ 
\ion{Ca}{1}  & 6455.600 & 2.520 & -1.370 & $ 46.3 \pm  3.7$ & 	  \ion{Co}{1}  & 5301.040 & 1.710 & -1.930 & $ 18.4 \pm  5.2$ \\ 
\ion{Ca}{1}  & 6572.800 & 0.000 & -4.280 & $ 25.5 \pm  3.8$ & 	  \ion{Co}{1}  & 5483.360 & 1.710 & -1.220 & $ 47.9 \pm  4.3$ \\ 
  &  &  &  &  & 						  \ion{Co}{1}  & 6093.150 & 1.740 & -2.340 & $ 13.9 \pm  4.1$ \\ 
\ion{Sc}{2} & 5239.820 & 1.450 & -0.760 & $ 69.2 \pm  5.5$ & 	    &  &  &  &   \\
\ion{Sc}{2} & 5318.360 & 1.360 & -1.700 & $ 19.7 \pm  5.1$ & 	  \ion{Ni}{1}  & 5082.350 & 3.658 & -0.590 & $ 85.4 \pm  6.9$ \\ 
\ion{Sc}{2} & 5526.820 & 1.770 &  0.150 & $ 83.8 \pm  4.2$ & 	  \ion{Ni}{1}  & 5088.540 & 3.850 & -1.040 & $ 28.2 \pm  6.9$ \\ 
\ion{Sc}{2} & 6245.620 & 1.510 & -1.040 & $ 44.5 \pm  3.6$ & 	  \ion{Ni}{1}  & 5094.420 & 3.833 & -1.070 & $ 20.9 \pm  6.9$ \\ 
\ion{Sc}{2} & 6300.690 & 1.510 & -1.960 & $  9.9 \pm  3.6$ & 	  \ion{Ni}{1}  & 5578.720 & 1.680 & -2.650 & $ 64.3 \pm  4.2$ \\ 
\ion{Sc}{2} & 6320.840 & 1.500 & -1.840 & $ 12.6 \pm  3.6$ & 	  \ion{Ni}{1}  & 5587.860 & 1.930 & -2.380 & $ 43.8 \pm  4.3$ \\ 
\ion{Sc}{2} & 6604.600 & 1.360 & -1.160 & $ 47.8 \pm  3.9$ & 	  \ion{Ni}{1}  & 5682.200 & 4.100 & -0.390 & $ 59.1 \pm  4.3$ \\ 
  &  &  &  &  & 						  \ion{Ni}{1}  & 5694.990 & 4.090 & -0.600 & $ 54.8 \pm  4.3$ \\ 
\ion{Ti}{1} & 5024.850 & 0.818 & -0.560 & $ 52.9 \pm  7.0$ & 	  \ion{Ni}{1}  & 5805.220 & 4.170 & -0.580 & $ 41.9 \pm  4.4$ \\ 
\ion{Ti}{1} & 5219.710 & 0.021 & -2.240 & $ 20.9 \pm  5.7$ & 	  \ion{Ni}{1}  & 5847.000 & 1.680 & -3.410 & $ 26.4 \pm  4.4$ \\ 
\ion{Ti}{1} & 5490.150 & 1.460 & -0.980 & $ 23.9 \pm  4.2$ & 	  \ion{Ni}{1}  & 6086.280 & 4.260 & -0.440 & $ 46.3 \pm  4.1$ \\ 
\ion{Ti}{1} & 5866.460 & 1.070 & -0.840 & $ 36.9 \pm  4.4$ & 	  \ion{Ni}{1}  & 6111.070 & 4.090 & -0.800 & $ 47.9 \pm  4.0$ \\ 
\ion{Ti}{1} & 6258.110 & 1.440 & -0.440 & $ 47.6 \pm  3.6$ & 	  \ion{Ni}{1}  & 6128.980 & 1.680 & -3.370 & $ 22.2 \pm  3.9$ \\ 
\ion{Ti}{1} & 6261.110 & 1.430 & -0.490 & $ 38.4 \pm  3.6$ & 	  \ion{Ni}{1}  & 6130.140 & 4.260 & -0.950 & $ 32.8 \pm  3.9$ \\ 
\ion{Ti}{1} & 6312.240 & 1.460 & -1.580 & $  7.6 \pm  3.6$ & 	  \ion{Ni}{1}  & 6175.370 & 4.089 & -0.550 & $ 56.0 \pm  3.7$ \\ 
  &  &  &  &  & 						  \ion{Ni}{1}  & 6176.820 & 4.088 & -0.260 & $ 74.0 \pm  3.7$ \\ 
\ion{V}{1} & 5668.370 & 1.080 & -1.000 & $  5.5 \pm  4.3$ & 	  \ion{Ni}{1}  & 6177.250 & 1.826 & -3.510 & $  9.4 \pm  3.7$ \\ 
\ion{V}{1} & 5670.850 & 1.080 & -0.460 & $  9.8 \pm  4.3$ & 	  \ion{Ni}{1}  & 6204.610 & 4.088 & -1.110 & $ 28.7 \pm  3.6$ \\ 
\ion{V}{1} & 5727.050 & 1.080 & -0.000 & $ 39.0 \pm  4.3$ & 	  \ion{Ni}{1}  & 6378.260 & 4.154 & -0.830 & $ 46.4 \pm  3.7$ \\ 
\ion{V}{1} & 5727.660 & 1.050 & -0.890 & $ 15.7 \pm  4.3$ & 	  \ion{Ni}{1}  & 6643.640 & 1.676 & -2.030 & $ 87.9 \pm  3.9$ \\ 
\ion{V}{1} & 5737.070 & 1.060 & -0.770 & $  7.3 \pm  4.4$ & 	  \ion{Ni}{1}  & 6772.320 & 3.658 & -0.970 & $ 51.3 \pm  3.8$ \\ 
\ion{V}{1} & 6111.650 & 1.043 & -0.710 & $ 13.1 \pm  4.0$ & 	  \ion{Ni}{1}  & 7748.890 & 3.700 & -0.380 & $ 87.7 \pm  2.8$ \\ 
\ion{V}{1} & 6216.350 & 0.280 & -0.900 & $ 25.4 \pm  3.6$ & 	  \ion{Ni}{1}  & 7797.590 & 3.900 & -0.350 & $ 73.6 \pm  2.8$ \\ 
\ion{V}{1} & 6251.830 & 0.287 & -1.340 & $ 22.9 \pm  3.6$ & 	    &  &  &  &   \\
  &  &  &  &  & 						  \ion{Zn}{1}  & 4722.160 & 4.030 & -0.370 & $ 66.9 \pm  9.4$ \\ 
\ion{Cr}{1}  & 5304.180 & 3.460 & -0.680 & $ 14.3 \pm  5.2$ & 	  \ion{Zn}{1}  & 4810.537 & 4.080 & -0.130 & $ 82.3 \pm  8.4$ \\ 
\ion{Cr}{1}  & 5312.860 & 3.450 & -0.580 & $ 13.4 \pm  5.1$ & 	  \ion{Zn}{1}  & 6362.350 & 5.790 &  0.140 & $ 28.7 \pm  3.7$ \\ 
\ion{Cr}{1}  & 5318.770 & 3.440 & -0.710 & $ 10.1 \pm  5.1$ & 	    &  &  &  &   \\
\ion{Cr}{1}  & 5480.510 & 3.500 & -0.830 & $ 10.2 \pm  4.3$ & 	  \ion{Sr}{1}  & 4607.340 & 0.000 &  1.906 & $ 41.4 \pm 10.8$ \\ 
\ion{Cr}{1}  & 5574.390 & 4.450 & -0.480 & $  2.7 \pm  4.2$ & 	    &  &  &  &   \\
\ion{Cr}{1}  & 5783.070 & 3.320 & -0.400 & $ 37.4 \pm  4.4$ & 	  \ion{Y}{2}  & 4900.120 & 1.030 & -0.090 & $ 56.5 \pm  7.4$ \\ 
\ion{Cr}{1}  & 5783.870 & 3.320 & -0.150 & $ 31.5 \pm  4.4$ & 	  \ion{Y}{2}  & 5087.430 & 1.084 & -0.160 & $ 50.0 \pm  6.9$ \\ 
\ion{Cr}{1}  & 5787.920 & 3.320 & -0.110 & $ 42.0 \pm  4.4$ & 	  \ion{Y}{2}  & 5200.420 & 0.992 & -0.570 & $ 51.9 \pm  5.8$ \\ 
\ion{Cr}{1}  & 6330.100 & 0.940 & -2.900 & $ 34.3 \pm  3.6$ & 	  \ion{Y}{2}  & 5402.780 & 1.839 & -0.440 & $ 15.2 \pm  4.7$ \\ 
  &  &  &  &  & 						    &  &  &  &   \\
\ion{Cr}{2} & 5305.870 & 3.830 & -1.970 & $ 39.9 \pm  5.2$ & 	  \ion{Ba}{2}  & 5853.690 & 0.604 & -0.910 & $ 78.7 \pm  4.4$ \\ 
  &  &  &  &  & 						  \ion{Ba}{2}  & 6141.730 & 0.704 & -0.030 & $134.9 \pm  3.9$ \\ 
\ion{Mn}{1} & 4265.920 & 2.940 & -0.440 & $ 55.8 \pm 13.5$ & 	  \ion{Ba}{2}  & 6496.910 & 0.604 & -0.410 & $112.6 \pm  3.7$ \\ 
\ion{Mn}{1} & 4470.130 & 2.940 & -0.550 & $ 56.3 \pm 12.4$ & & & & & \\
\enddata
\end{deluxetable}

\end{document}